\documentclass{ws-ijseke}
\usepackage{array}
\usepackage{mathrsfs}
\usepackage{algorithm}
\usepackage{amsmath}
\usepackage{graphicx}
\usepackage{setspace}
\usepackage{lineno}
\usepackage{color}
\usepackage{enumerate}
\usepackage{enumitem}
\usepackage{hyperref}

\usepackage{float}
\usepackage{breakurl}
\begin{document}

\markboth{Miroslav Bures, Bestoun S. Ahmed, Kamal Z. Zamli}
{Prioritized Process Test: An Alternative to Current Process
Testing Strategies}

%
\catchline{01}{01}{2003}{}{}
%

\title{Prioritized Process Test: An Alternative to Current Process Testing Strategies\footnote{Preprint of an article submitted for consideration in the Int'l Journal of Software Engineering and Knowledge Engineering }}

\author{Miroslav Bures}

\address{Software Testing Intelligent Lab (STILL), Department of Computer
Science, Faculty of Electrical Engineering Czech Technical University,
Karlovo nam. 13, 121 35 Praha 2 \\ Czech Republic
\email{buresm3@fel.cvut.cz}
}

\author{Bestoun S. Ahmed}

\address{Department of Mathematics and Computer Science, Karlstad University, Sweden\\
Software Testing Intelligent Lab (STILL), Department of Computer
Science, Faculty of Electrical Engineering Czech Technical University,
Karlovo nam. 13, 121 35 Praha 2 \\ Czech Republic\\
\email{bestoun@kau.se}
}

\author{Kamal Z. Zamli}

\address{IBM Center of Excellence, Faculty of Computer Systems and Software
Engineering\\ University Malaysia Pahang, Gambang, Malaysia\\
\email{kamalz@ump.edu.my}
}

\maketitle

\begin{history}
\end{history}

\begin{abstract}
Testing processes and workflows in information and Internet of Things systems is a major part of the typical software testing effort. Consistent and efficient path-based test cases are desired to support these tests. Because certain parts of software system workflows have a higher business priority than others, this fact has to be involved in the generation of test cases. In this paper, we propose a Prioritized Process Test (PPT), which is a model-based test case generation algorithm that represents an alternative to currently established algorithms that use directed graphs and test requirements to model the system under test. The PPT accepts a directed multigraph as a model to express priorities, and edge weights are used instead of test requirements. To determine the test-coverage level of test cases, a test-depth-level concept is used. We compared the presented PPT with five alternatives (i.e., the Process Cycle Test, a naive reduction of test set created by the Process Cycle Test, Brute Force algorithm, Set-covering Based Solution and Matching-based Prefix Graph Solution) for edge coverage and edge-pair coverage. To assess the optimality of the path-based test cases produced by these strategies, we used fourteen metrics based on the properties of these test cases and 59 models that were created for three real-world systems. For all edge coverage, the PPT produced more optimal test cases than the alternatives in terms of the majority of the metrics. For edge-pair coverage, the PPT strategy yielded similar results to those of the alternatives. Thus, the PPT strategy is an applicable alternative, as it reflects both the required test coverage level and the business priority in parallel.
\end{abstract}


\section{Introduction}

\begin{figure*}
\begin{centering}
\includegraphics[width=13cm]{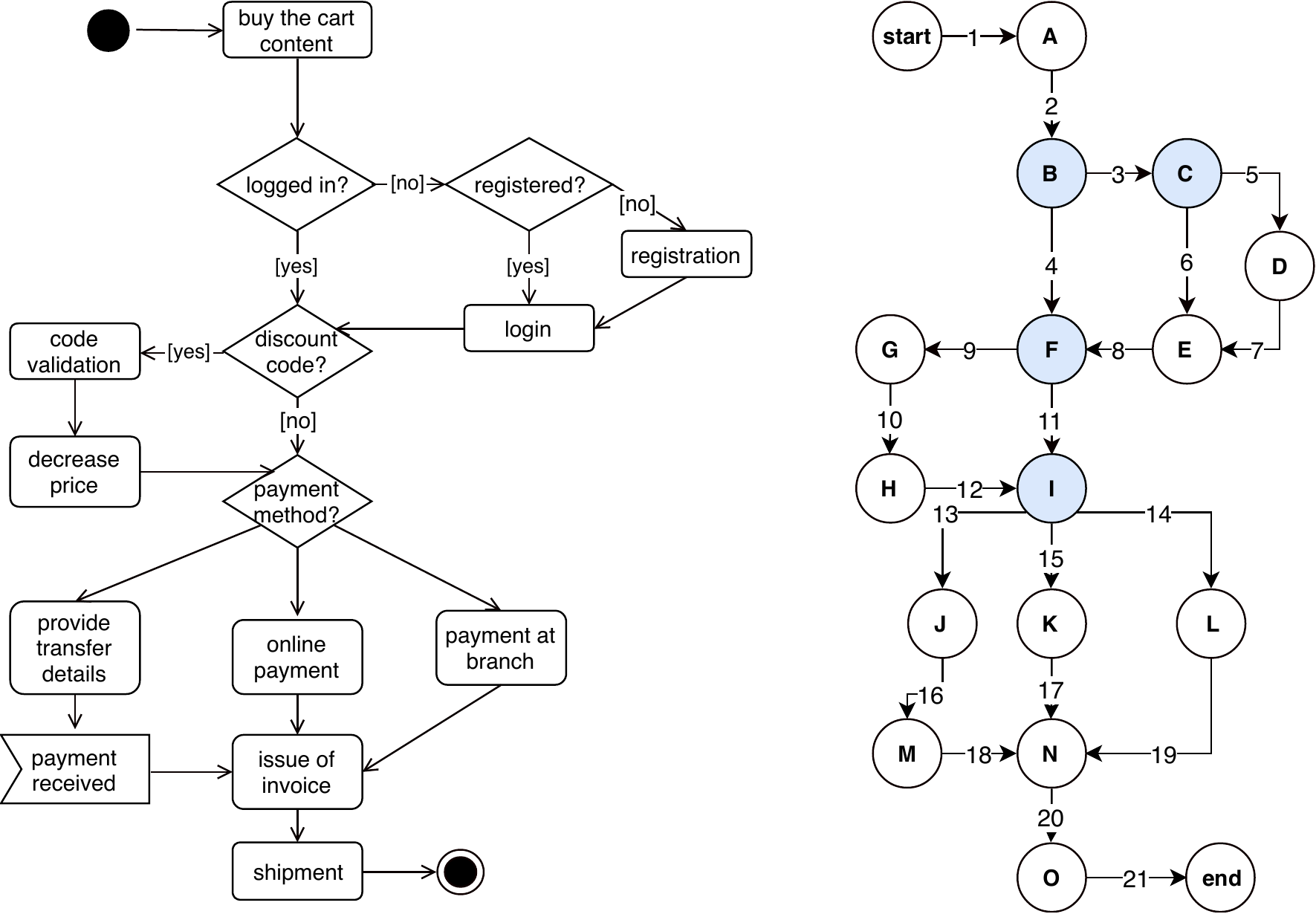}
\par\end{centering}
\caption{\label{fig:Coversion-of-UML}An example of UML Activity Diagram conversion
to a directed graph as a SUT model}
\end{figure*}

In current software and Internet of Things (IoT) systems, the testing of processes and workflows is one of the major testing techniques \cite{offutt2008introduction,myers2011art}. This type of testing also represents a considerable part of the overall software testing budget; for instance, Eldh et al. estimates that these costs are between 40\% and 80\% of the total software development project costs \cite{Eldh:2006}. The efficiency of process tests (i.e., tests that are based on a flow of actions in a system under test (SUT) to detect possible defects in SUT processes \cite{koomen2013tmap}) strongly depends on the quality and consistency of the created test cases \cite{utting2010practical,myers2011art,gupta2008approach}. Hence, the generation of consistent and efficient path-based test cases (i.e., test cases that are based on flows or logical paths that can be executed in an SUT) \cite{offutt2008introduction} has been a subject of interest of the Model-based Testing discipline in the last decade \cite{schieferdecker2012model,utting2012taxonomy}. For test case generation, an underlying SUT model is needed. For path-based test cases, this model is based on a directed graph \cite{offutt2008introduction,utting2010practical}. A Unified Modeling Language (UML) activity diagram, which is the current established design option for the modeling of workflows and processes in information systems \cite{budgen2011empirical,rumpe2016modeling} can be converted into a directed graph \cite{utting2010practical,offutt2008introduction} (a simple example is presented in Fig \ref{fig:Coversion-of-UML}). A process modeling alternative to UML is Business Process Model Notation (BPMN), for which a natural solution is a conversion to a Petri net \cite{wieczorek2009applying,kalenkova2017process}; however, for model-based testing purposes, a conversion to a directed graph is also feasible \cite{mishra2015graph}. Several strategies were proposed to generate path-based test cases in the literature; for example \cite{yoo2012regression,dwarakanath2014minimum,li2012better,gokcce2006coverage,belli2007coverage,panthi2015generating}.

In this paper, we present the Prioritized Process Test (PPT), which is an alternative approach to path-based test case generation from an SUT model based on a directed graph. This approach enables good flexibility for different testing goals, allows the test coverage level of the test cases to be scaled, and in parallel with this scaling, it provides a prioritization mechanism to optimize the test set (additional details are provided in Section \ref{sec:Prioritized-Process-Test}). In the experiments, we compare this strategy with five alternatives to evaluate its performance and the properties of the produced test set.

The rationale behind this study is twofold. First, due to its the importance, the process-based approach and workflow testing is worth further exploring to find an alternative to previously published strategies. Second, defining exact optimization criteria is a challenging task. As we can see in Section \ref{subsec:Test-set-optimization}, various aspects can be further used to optimize the test sets.

The motivation for this approach is underpinned by several points. As we explain in detail in Section \ref{subsec:Motivation-for-an}, algorithms that are built upon an SUT model that is based on a directed graph and a set of test requirements (explained in Section \ref{subsec:Prioritization-of-the}), which are present in the test cases, may reach their limit when trying to satisfy a defined test coverage level together with the prioritization of the SUT model parts, which are visited by the test cases. Test requirements practically allow only two levels of priority (i.e., priority and not priority), whereas in practical test prioritization, more levels are usually used \cite{achimugu2014systematic,van2013tmap}. Another limit of algorithms based on established test requirements is the possible limit of visiting priority paths sideways in the processes that are captured by the SUT model. Although this limit can be solved by employing multiple sets of test requirements, the number of these alternative combinations can grow to an extent that makes it challenging to keep the SUT model actual with the SUT. Finally, for commonly used process models, algorithms that accept only a directed graph as an input require conversion of the parallel edges in the model into additional nodes, which leads to more extensive models that may be impractical to maintain. Hence, the objective of this paper is to propose an alternative approach to prioritized path-based test generation, which mitigates the aftermentioned issues and compare it with the current alternatives.

The contributions of the paper are as follows:
\begin{enumerate}
\item the modeling approach for path-based testing is discussed and an alternative problem model is proposed, which, in our opinion, better suits the needs of the current software testing industry; 
\item the PPT test case generation approach, which employs this alternative model and aims to optimize the test set, is proposed to reflect the business priorities defined in the model; 
\item fourteen evaluation criteria are proposed to evaluate the optimality of the generated test case; and 
\item experimental results from 59 problem instances created from three real-world systems are presented to compare the PPT with five alternatives, which provides insight into the performance of the PPT and the five compared algorithms from the viewpoint of the presented evaluation criteria. 
\end{enumerate}

This paper is organized as follows. In Section \ref{sec:Problem-Definition}, we define the problem and introduce our motivation to develop an alternative strategy; we present this strategy in Section \ref{sec:Prioritized-Process-Test}. Section \ref{sec:Experiments} shows the experimental methodology for evaluating the proposed strategy, the results and the discussion. We analyze and present some possible threats to validity in Section \ref{sec:Threats-to-Validity}. In Section \ref{sec:Related-Work}, we summarize and discuss related works. The last section concludes the paper.

\section{\label{sec:Problem-Definition}Preliminaries and Problem Definition }

For path-based testing, the SUT model is commonly defined as a directed graph $G=$($D$, $A$), where $D$ is a set of nodes, $D\neq\emptyset$, and $A$ is a set of edges \cite{offutt2008introduction}. $A$ is a subset of $D\times D$. In the model, one start (initial) node $d_{s}\in D$ is defined. $D_{e}\subseteq D$ is a set of end nodes, $D_{e}\neq\emptyset$. 

The nodes in the graph can be used to model decision points in the process and the SUT actions or functions, which are executed between particular decision points. Alternatively, nodes can be used to model decision points in the process and edges can be an abstraction of either:
\begin{enumerate}
\item one physical step performed in the SUT, or
\item a linear sequence of physical steps in the SUT without the possibility
to select an alternative way (i.e., no decision point is implemented
in this sequence of physical steps).
\end{enumerate}
In this paper, we use the SUT model presented in
Fig. \ref{fig:Coversion-of-UML} as a running example to document
the presented concepts and algorithms. In our example, $D_{R}=\{start, A, B, C, D, E, F, G, H, I, J, K, L, M, N, O, end \}$ and $A_{R}=\{1, 2, 3, 4, 5, 6, 7, 8, 9, 10, 11, 12, 13, 14, 15, 16, 17, 18, 19, 20 \}$.

The granularity of the physical steps can differ by the level of abstraction for which the test cases are prepared. For example, for the design of a business end-to-end test, this step typically corresponds to the SUT function defined on the conceptual level of SUT design.

The test case $t$ is a sequence of nodes $d_{1},d_{2},..,d_{n}$, with a sequence of edges $a_{1},a{}_{2},..,a_{n-1}$, where $a_{i}=(d_{i,}d_{i+1})$, $a_{i}\in A$, $d_{i}\in D$. Moreover, $d_{1}=d_{s}$ and $d_{n}\in D_{e}$. When these conditions are satisfied, we can alternatively denote the test case as a sequence of nodes $d_{1},d_{2},..,d_{n}$ or edges $a_{1},a_{2},..,a_{n-1}$ only. We refer to the individual nodes and edges in the test case $t$ as test case steps. The test set $T$ is a set of test cases. Alternatively, some studies refer to test cases as test paths \cite{li2012better,offutt2008introduction}. In our example, a sequence such as $\{$start, 1, A, 2, B, 4, F, 11, I, 15, K, 17, N, 19, O, 20, end$\}$ can be a test case.

\subsection{\label{subsec:Test-coverage-criteria}Test Coverage Criteria}

The extent of a test set $T$ is determined by coverage criteria. First, let us describe the problem without considering the prioritization of SUT functions. 

A trivial coverage criterion example would be following each edge $a\in A$ minimally once in the test set $T$. Usually, this criterion is referred to as \textit{All Edge Coverage}. Alternatively, \textit{All Node Coverage} can also be defined, which requires each node $d\in D$ to appear minimally once in the test set $T$. In fact\textit{, All Node Coverage} is even weaker than \textit{All Edge Coverage}. \textit{All Edge Coverage} serves for lightweight testing that is usually suitable for smoke tests (i.e., tests detecting that the essential functionality of the SUT has not been disrupted by defects) or routine regression tests (i.e., repetitive tests following SUT updates to determine if new defects are present in parts of the SUT, which were considered free of defects) \cite{koomen2013tmap}. For more thorough workflow testing, a higher level of test coverage is usually needed.

An opposite extreme to \textit{All Edge Coverage} (in terms of the extent of $T$) that exercises all possible paths in $G$, starting with $d_{s}$ and ending with any of $d_{e}\in D_{e}$, is usually referred to as \textit{All Paths Coverage} \cite{offutt2008introduction}. However, such tests would be too extensive and result in high demand for resources and time for testing. Thus, a compromise is needed to determine a practical level of test coverage. 

One possibility is the \textit{Test Depth Level (TDL)} criterion used in the Process Cycle Test (PCT) technique, as defined in TMap \textit{Next} \cite{koomen2013tmap}. PCT uses $G$ as the SUT model and produces test cases $t$ (as defined above). The TDL criterion is defined as:

\begin{enumerate} 

\item $TDL=1$ if $\forall a\in A$, edge $a$ occurs at least once in at least one test case $t\in T$ (which is equivalent to \textit{All Edge Coverage}). 

\item $TDL=n$ if the following conditions are satisfied: for each node $d\in D$, the $S_{d}$ is a set of all possible paths in $G$ starting
with an edge incoming to the node $d$ and, followed by a sequence
of $n-1$ edges outgoing from node $d$. Then, $\forall d\in D$,
the test cases of the test set $T$ contain all paths from $S_{d}$. 

\end{enumerate}

Alternatively, $TDL=2$ is referred to as \textit{Edge-pair Coverage}. In practice, $TDL>3$ is not commonly used for the generation of test cases \cite{van2013tmap,koomen2013tmap}.

Another concept used in the area is \textit{Prime Path Coverage} (PPC), which eliminates redundancy in the created tests \cite{offutt2008introduction}. To satisfy the PPC criterion, each reachable prime path in $G$ has to be a subpath of a test case $t\in T$. A path $p$ from $d_{1}$ to $d_{2}$ is prime if the following conditions are satisfied:
\begin{enumerate}
\item $p$ is simple, which means no node $d\in D$ appears more than once in $p$ (i.e., $p$ does not contain any loops). The only exceptions are $d_{1}$ and $d_{2}$, which can be identical ($p$ itself can be a loop).
\item $p$ is not a sub-path of any other simple path in $G$.
\end{enumerate}

By its nature, PPC tends to produce more extensive test cases with high coverage. In this sense, the advantage of TDL is the possibility to flexibly set the test coverage level. It is worth noting that in the graph $G$ used as a SUT model, other coverage criteria can also be defined (e.g., all simple paths, all simple round trips or all complete round trips), but they are out of the scope of this paper. Importance or priorities of individual functions of the SUT are reflected neither in TDL nor Prime Path Coverage concepts. 

To illustrate some of the explained coverage criteria using the running
example, \textit{All Edge Coverage} (equivalent to $TDL=1$) and \textit{All
Node Coverage} are satisfied by a $T_{R1}=\{t_{1},t_{2},t_{3}\}$,
where $t_{1}=\{$start, 1, A, 2, B, 3, C, 5, D, 7, E, 8, F, 9, G, 10, H ,1 2, I, 13, J, 16, M, 18, N, 20, O, 21, end$\}$, $t_{2}=\{$start, 1, A, 2, B, 4, F, 11, I, 14, L, 19, N, 20, O, 21, end$\}$, and
$t_{3}=\{$start, 1, A, 2, B, 3, C, 6, E, 8, F, 9, G, 10, H, 12, I, 15, K, 17, N, 20, O, 21, end$\}$.

For instance, \textit{Edge-pair Coverage} (equivalent to $TDL=2$)
is satisfied by a $T_{R2}=\{t_{1},t_{2},t_{3},t_{4},$ $t_{5},t_{6}\}$,
where $t_{1}=\{$start, 1, A, 2, B, 3, C, 5, D, 7, E, 8, F, 9, G, 10, H, 12, I, 13, J, 16, M, 18, N, 20, O, 21, end$\}$, $t_{2}=\{$start,  1, A, 2, B, 4, F, 9, G, 10, H, 12, I, 14, L, 19, N, 20, O, 21, end$\}$, $t_{3}=\{$start, 1, A, 2, B, 3, C, 6, E, 8, F, 11, I, 13, J, 16, M, 18, N, 20, O, 21, end$\}$, $t_{4}=\{$start, 1, A, 2, B, 4, F, 11, I, 14, L, 19, N, 20, O, 21, end$\}$, $t_{5}=\{$start, 1, A, 2, B, 4, F, 11, I, 15, K, 17, N, 20, O, 21, end$\}$, and $t_{6}=\{$start, 1, A, 2, B, 4, F, 9, G, 10, H, 12, I, 15, K, 17, N, 20, O, 21, end$\}$.  The edge pairs, which have to be present in at least one of the test cases of $T$ to satisfy the \textit{Edge-pair Coverage} are, for example, {2-3} and {2-4 }for { B}, {4-9}, {4-11}, {8-9} and {8-11 }for {F}, {11-13}, {11-14}, {11-15}, {12-13}, {12-14}, and {12-15 }for { I} and so forth.

\subsection{\label{subsec:Prioritization-of-the}Prioritization of the SUT Functions}

To achieve more efficient test cases that address real business priorities in the testing process, particular SUT functions (nodes $D$ or edges $A$) should be prioritized. For prioritization, we can use the \textit{Test Requirements} \cite{dwarakanath2014minimum,li2012better,offutt2008introduction}.

The particular form of a test requirement depends on the selected test coverage criteria. For Edge Coverage, a test requirement is an edge $a\in A$, which has to be present in at least one test case $t\in T$. Thus, in contrast to \textit{All Edge Coverage}, we do not require all edges of $G$ to be covered, and cover only those defined by a set of test requirements. By analogy, for Node Coverage, a test requirement is a node $d\in D$ that must be present in at least one test case $t\in T$.

Generally, when we compose the test case as a sequence of nodes $d_{1},d_{2},..,d_{n}$ with a sequence of edges $a_{1},a{}_{2},..,a_{n-1}$, where $a_{i}=(d_{i,}d_{i+1})$, $a_{i}\in A$, $d_{i}\in D$, a test requirement is a path in $G$ that must be a sub-path of at least one test case $t\in T$ \cite{li2012better} (in some studies, the test requirement is discussed as a node or an edge to be toured by at least one test case $t\in T$ \cite{offutt2008introduction}; however, in this paper, we consider the test requirement as defined previously). We denote the set of test requirements as $R$. 

An adequate set of test requirements can express priority edges or nodes (composing the priority parts of the SUT workflows). If an edge $a\in A$ is a priority edge and $a$ is not adjacent to any other priority edges in $G$, a test requirement $\{a\}$ is added to $R$. If a node $d\in D$ is a priority node and its incoming and outgoing edges are non-priority edges only, a test requirement $\{d\}$ is added to $R$. If a path $p$ in $G$ consists of priority edges and priority nodes only and $p$ is not a sub-path of any other path in $G$, which also consists of priority edges and priority nodes only, a test requirement $\{p\}$ is added to $R$. However, in a number of published algorithms, the test requirements are also used to determine the test coverage criteria, which could limit the usage of test requirements for prioritization. We analyze this issue in Section \ref{subsec:Motivation-for-an}.

\subsection{\label{subsec:Test-set-optimization}Test Set Optimization Criteria}

In the optimization of the test set $T$, a number of criteria can be used; for instance:

\begin{enumerate}
\item $\sum_{t\in T}\mid t\mid$, the total number of test case steps (nodes
or edges) in the test case $t$. The lower total number of test steps
would be the lower we can expect the testing costs in the phases of
detailed test design and test execution. In the running example used
in this paper, $\sum_{t\in T_{R1}}\mid t\mid=71$ and $\sum_{t\in T_{R2}}\mid t\mid=128$.
\item A total number of nodes can be found as an alternative criterion to
a total number of test case steps \cite{li2012better}. The nodes
in the test case usually require entering the test data, which makes
the testing process costlier. However, this criterion is principally
very similar to the total number of test case steps. In the running
example, the total number of nodes is 37 for $T_{R1}$ and 67 for
$T_{R2}$. 
\item $\mid T\mid$ is the total number of the test cases. A higher number
of test cases can imply higher maintenance of the test set. However,
end-to-end test cases with test paths that are too long could also
be counterproductive. For instance, when a defect in the SUT disables
execution of the test case and we are waiting to have this defect
fixed, the rest of the test cases cannot be executed. The longer the
test cases are, the larger the extent of the SUT that can become unavailable
to test. Waiting for defect fixes is a classical situation software
testers experience daily. In the running example, $\mid T_{R1}\mid=3$
and $\mid T_{R2}\mid=6$.
\item $\frac{\mid R\mid}{\mid T\mid}$ which express how many test requirements
are covered by one test case \cite{li2012better}. At first glance,
the higher this ratio is, the closer the test set $T$ is to the optimum.
However, consider the following situation: a test requirement $r\in R$
cannot be feasible for some reason, and we do not have this information
during the high-level abstraction of the test design. Thus, $r$ blocks
the test case from completion. If this test case contains another
test requirement, it will be blocked. For this reason, we might prefer
to keep the $\frac{\mid R\mid}{\mid T\mid}$ ratio lower.
\end{enumerate}

For the practical testing process, determination of the best optimization criteria could be challenging. Individual algorithms for generating path-based test cases from to generate paths-based test cases from G can perform differently in satisfying these individual optimization criteria.

\subsection{\label{subsec:Motivation-for-an}Motivation for an Alternative Approach}

There are several factors that motivate us to develop an alternative to the current path-based strategies that use a directed graph $G$ and a set of test requirements $R$ as an SUT model.

(1) The test requirements $R$, which have been extensively used in previous work, can be used to define either the prioritized parts of the SUT model that should be examined in the test cases or the general test coverage level of the test set $T$ (e.g., Prime Path coverage \cite{offuttoolsapplication}). Unfortunately, when test requirements are used to specify the general test coverage criteria, they cannot also be used to determine which parts of the SUT should be considered a priority to be covered by the tests, along with the general test coverage criterion, which limits many published algorithms that use the test requirement concept, such as the Set-Covering Based Solution algorithm or the Matching-Based Prefix Graph Solution \cite{li2012better}. For example, we can use \textit{Edge-Pair Coverage} as the test coverage level (explained previously in Section \ref{subsec:Test-coverage-criteria}). For \textit{Edge-Pair Coverage,} $R$ contains each possible pair of adjacent edges in $G$ that shall be present in $T$. In such a
situation, we cannot use $R$ further to specify, which parts of the SUT model $G$ are considered as priorities. 

(2) The test requirements can be used to determine which parts of
the SUT should be prioritized in only two priority levels (i.e., which
parts of the model, as captured by a set of test requirements, should
be prioritized, and which parts should not be prioritized). In contrast,
in most software engineering and management practices, more priority
levels (typically three) are used \cite{achimugu2014systematic,van2013tmap}.
A trivial solution to this issue involves transforming the priority
levels into a set of test requirements. However, during such a transformation,
information about priority levels is reduced to two states: priority
and nonpriority. Therefore, more detail about the priority levels
is not available to an algorithm that accepts $G$ and $R$ as inputs.
Capturing more priority levels in the SUT model would allow the formulation
of algorithms that more efficiently optimize the test according to
these priorities. 

(3) When a path $p$ in an SUT model $G$ is considered a priority scenario in the SUT process, we are primarily interested in examining this process in the test cases. However, for crucial SUT processes, it might also be useful to exercise the sideways of $p$, which typically represents parts of the workflow that can be initiated from the path $p$. By sideway in this context, we mean a path $p_{2}$ that starts with a nonempty intersection with path $p$. Such a situation can be modeled using several sets of test requirements. Let $R_{1}$, $R_{2}$ and $R_{3}$ be three sets of test requirements. In $R_{1}$, we capture the path $p$. In $R_{2}$, we capture only the immediate sideways of $p$. For intense testing, we can define an $R_{3}$ that captures another possible sideways of $p$. Practically, this approach requires the maintenance of multiple sets of test requirements, which merges three aspects in parallel: (a) the required level of test coverage, (b) the prioritized parts of the SUT model and (c) the sideways we need to test. By considering a simple real-life example in which the test analyst would be interested in generating test cases for three test coverage levels using three priority levels (i.e., low, medium and high) and two sideways levels, we end up with 18 possible sets of test requirements, which must be maintained and representative of the actual SUT. Such modeling is possible but demanding for the analyst, and the resulting model can be prone to creation and maintenance errors. A strategy that uses only the prioritized parts of the SUT to generate test cases and systematically examines the paths of the workflow that are initiated from these prioritized parts shall beprovided. 

Another issue occurs as a result of using a directed graph in SUT modeling:

(4) If a directed graph is used, parallel edges (i.e., edges that have the same start and end nodes) are not allowed in the model. In numerous models based on UML Activity Diagrams, parallel edges are used frequently. This situation can also be modeled using a directed graph, where special nodes must be used to distinguish the parallel edges. However, such a situation can lead to more complex models, which are difficult to maintain and result in a greater amount of data that must be processed by the test case generation algorithms. To illustrate this issue, consider a SUT model that is captured as a directed multigraph that consists of 15 nodes and 40 edges, 12 of which are parallel to other edges of the graph. To model the problem with $G$, we need to add 12 more nodes and 12 more edges so that the final model consist of 27 nodes and 52 edges (note that the number of nodes almost doubled).

The second, third, and fourth issues do not limit the current concept of SUT modeling based on $G$ and $R$. With a set of data transformations, the problem can be converted to the form of $G$ and $R$ and current algorithms can be used, as explained in the discussion above. However, to solve the first issue, an alternative approach must be proposed.

\section{\label{sec:Prioritized-Process-Test}Prioritized Process Test}

In the proposed prioritized process test (PPT), we combine the TDL criterion (see Section \ref{subsec:Test-coverage-criteria}) with the priorities of the SUT functions based on the weights of the G edges. The PPT technique generates test cases that focus on covering the prioritized parts of the workflows using more test steps and deliberately covers nonpriority parts with a lower number of steps. In this section, we first describe the extended SUT model, which includes the prioritization of the SUT functions. Then, we define additional coverage criteria for the PPT technique, i.e., Prioritized Test Level. Finally, we provide details of the PPT strategy for the generation of the test set $T$.

\subsection{\label{subsec:Prioritization-of-SUT}Prioritization of SUT Functions
and Coverage Criteria}

PPT uses the SUT model defined as a weighted multigraph $\mathfrak{\mathcal{G}}=(D,A,s,t)$, where $D$ is a set of nodes, $D\neq\emptyset$, and $A$ is a set of edges. Here, $s:\,D\rightarrow A$ assigns each edge to its source node and $t:\,D\rightarrow A$ assigns each edge to its target node. One start node $d_{s}\in D$ is defined. The set $D_{e}\subseteq D$ contains the end nodes of $\mathfrak{\mathcal{G}}$, ${D_{e}\neq\emptyset}$. For each edge ${d\in D}$, a priority $p$ is defined, $p\in\{high,medium,low\}$. When a priority is not defined, it is considered $low$. Then, $A_{h}$ is a set of high-priority edges, $A_{m}$ is a set of medium-priority edges, and $A_{l}$ is a set of low-priority edges, where $A_{h}\cup A_{m}\cup A_{l}=A$, $A_{h}\cap A{}_{m}=\emptyset$, $A_{m}\cap A_{l}=\emptyset$, $A_{h}\cap A_{l}=\emptyset$. When modeling the SUT, priorities are determined by test analysts. Various techniques and approaches can be used, such as the Product Risk Analysis (PRA) presented in the BDTM approach \cite{van2013tmap}. Other approaches can be found in the study by Achimugu \textit{et al.} \cite{achimugu2014systematic}.

To determine the test coverage of individual parts of the SUT, the Prioritized Process Test uses two concurrent coverage criteria: (1) the TDL criterion, as mentioned, and (2) Prioritized Test Level (PTL). PTL can be set to values ${high,medium}$ and is defined as:

\begin{enumerate}
\item $PTL=high$ if $\forall a\in A_{h}$, edge $a$ occurs at least once
in at least one test case $t\in T$. 
\item $PTL=medium$ if $\forall a\in A_{h}\cup A_{m}$, edge $a$ occurs
at least once in at least one test case $t\in T$. 
\end{enumerate}

To determine the test coverage, the TDL criterion is used as specified in Table \ref{tab:Specification-of-TDL}. Let $P$ be a set of paths in $\mathfrak{\mathcal{G}}$ such that $\forall p\in P$, $p$ must be a sub-path of a test $t\in T$ to satisfy the test coverage criteria. These paths have length 1 for $TDL=1$. In Table \ref{tab:Specification-of-TDL}, $P$ is specified for particular possible combinations of PTL and TDL.

\begin{table}
\begin{centering}
\caption{\label{tab:Specification-of-TDL}Specification of TDL by particular
value of PTL}
\par\end{centering}
\centering{}%

\begin{tabular}{|>{\centering}p{1.8cm}|>{\centering}p{3cm}|>{\centering}p{3cm}|}
\hline 
Coverage Criteria & $PTL=high$ & $PTL=medium$\tabularnewline
\hline 
\hline 
$TDL=1$ & $P=A_{h}$  & $P=A_{h}\cup A_{m}$ \tabularnewline
\hline 
$TDL=n$, $n>1$  & $P$ = set of all paths identified in $\mathfrak{\mathcal{G}}$ by
$TDL=n$ criterion, which start with any of $a\in A_{h}$ & $P$ = set of all paths identified in $\mathfrak{\mathcal{G}}$ by
$TDL=n$ criterion, which start with any of $a\in A_{h}\cup A_{m}$ \tabularnewline
\hline 

\end{tabular}
\end{table}

\subsection{\label{subsec:Algorithm-for-generation} Test Generation Strategy}

In this paper, we present the optimized version of the test generation strategy, the initial version of which we published in \cite{bures2017prioritized}. The version presented in this paper produces $T$ in a better runtime due to optimization of Algorithm \ref{alg:SelectRelevantTDLPaths(D,ALLTDL,TDL,PTL)} and the optimization of the physical implementation of the algorithm in the Oxygen platform used in the subsequent experiments. 

The strategy is composed of several algorithms. PPT test cases are generated by Algorithm \ref{alg:GenerateTestCases(G,TDL,PTL)}. Inputs to Algorithm \ref{alg:GenerateTestCases(G,TDL,PTL)} are the model of the SUT $\mathcal{G}$ and the selected TDL and PTL values. Output of the algorithm is the test set $T$. The test cases are specified as a sequence of $\mathscr{\mathcal{G}}$ edges $a_{1},a{}_{2},..,a_{n-1}$.

The main Algorithm \ref{alg:GenerateTestCases(G,TDL,PTL)} uses Algorithms \ref{alg:GetAllTDLPathsForNode(d,depth,ALL,S)}, \ref{alg:SelectRelevantTDLPaths(D,ALLTDL,TDL,PTL)} and \ref{alg:CreateTestCases(PTAB,ALLE2E)}. Subsequently, Algorithm \ref{alg:CreateTestCases(PTAB,ALLE2E)} uses Algorithms \ref{alg:SelectBestE2EPath(PTAB,ALLE2E)}, \ref{alg:RemoveUnnecessaryE2EPaths(PTAB,ALLE2E)} and \ref{alg:RemoveUsedTDLPaths(PTAB,b)}.

The principle of the PPT algorithm is the following. The main \textbf{Algorithm \ref{alg:GenerateTestCases(G,TDL,PTL)}}, which produces the test set $T$, starts with the initial identification of the paths specified by the $TDL$ criterion, which should be present in the test cases (the set $ALLTDL$, line 6). This identification is done by Algorithm \ref{alg:GetAllTDLPathsForNode(d,depth,ALL,S)}. Then, using Algorithm \ref{alg:SelectRelevantTDLPaths(D,ALLTDL,TDL,PTL)}, only paths starting with an edge of priority $high$ (or $high$ and $medium$, depending on the $PTL$ criterion) are filtered (set $P$, line 8). The set $P$ represents the paths that must be present in the test set $T$ to satisfy the coverage criteria. In the next step of Algorithm \ref{alg:GenerateTestCases(G,TDL,PTL)}, all possible end-to-end paths in $\mathcal{G}$, starting from the $d_{s}\in D$ and ending in any node $d_{e}\in D_{e}$ that contain a path $p\in P$ are identified (set $ALLE2E$, line 9). These end-to-end paths are candidates for the test cases of the test set $T$. However, in this phase, the set $ALLE2E$ is not optimal and is going to be reduced by Algorithm \ref{alg:CreateTestCases(PTAB,ALLE2E)} in the last step (line 13). 

In \textbf{Algorithm \ref{alg:CreateTestCases(PTAB,ALLE2E)}}, the optimization process is as follows. From $ALLE2E$, we select the end-to-end paths that contain most $p\in P$ using Algorithm \ref{alg:SelectBestE2EPath(PTAB,ALLE2E)} (line 3). Then, the end-to-end paths of $ALLE2E$ that are not needed because the particular $p\in P$ is already contained in the other end-to-end path are removed from the selection by Algorithm \ref{alg:RemoveUnnecessaryE2EPaths(PTAB,ALLE2E)} (line 4). During this process, paths $p\in P$ that have been already contained by some of the selected end-to-end paths are removed from further processing by Algorithm \ref{alg:RemoveUnnecessaryE2EPaths(PTAB,ALLE2E)} (line 5).

\begin{algorithm}

$T$  $\leftarrow$  $\emptyset$, $\quad$$P$ $\leftarrow$ $\emptyset$,
$\quad$$ALLTDL$ $\leftarrow$ $\emptyset$, $\quad$$ALLE2E$ $\leftarrow$
$\emptyset$

Set PTAB as empty

Initialize new empty stack S

depth $\leftarrow$ TDL 

\textbf{For} (each $d\in D$) \textbf{do}

$\quad$$ALLTDL$ $\leftarrow$ $ALLTDL$ $\cup$ GetAllTDLPathsForNode($d$,
depth, $ALLTDL$, S) 

\textbf{End for }

$P$ $\leftarrow$ SelectRelevantTDLPaths($D$, $ALLTDL$, TDL, PTL)

$ALLE2E$ $\leftarrow$ \{ $z$ \textbar{} $z$ is path in $\mathfrak{\mathcal{G}}$
starting with node $d_{s}\in D$ and ending with any node $d_{e}\in D_{e}$
and there exist a path $p\in P$ such that $p$ is a sub-path of $z$ \} 

\textbf{For} (each $p\in P$) \textbf{do}

$\quad$add $p$ to indexed table PTAB, $p$ is indexed by the second
node from $p$ 

\textbf{End for}

$T$ $\leftarrow$ CreateTestCases(PTAB, $ALLE2E$)

\caption{\label{alg:GenerateTestCases(G,TDL,PTL)}GenerateTestCases($\mathcal{G}$,
TDL, PTL) \textbf{$\qquad$ Output:} test set $T$}
 
\end{algorithm}

\begin{algorithm}
depth $\leftarrow$ depth-1

\textbf{If} (depth\textless 0) \textbf{then}

$\quad$Create a new path from a sequence of edges stored in stack
S and add it to $ALLTDL$

\textbf{End if}

$O$ $\leftarrow$ set of edges outgoing from $d$ 

\textbf{For} (each $o\in O$) \textbf{do} 

$\quad$Push $o$ to stack S

$\quad$$d_{o}$ $\leftarrow$ node at the end of edge $o$

$\quad$GetAllTDLPathsForNode($d_{o}$, depth, $ALLTDL$, S)

$\quad$remove $o$ from stack S 

\textbf{End for}

\textbf{If} (stack S is empty) \textbf{then} 

$\quad$return $ALLTDL$

\textbf{End if }

\caption{\label{alg:GetAllTDLPathsForNode(d,depth,ALL,S)}GetAllTDLPathsForNode($d$,
depth, $ALL$, S)$\qquad$\textbf{Output:} Iterative contribution
to $ALLTDL$ for node $d$}
\end{algorithm}

\begin{algorithm}
\textbf{For} (each $c\in ALLTDL$) \textbf{do}

$\quad$$e$ $\leftarrow$ the first edge of $c$

$\quad$\textbf{If} ((PTL=$high$ and priority of $e$ is $high$)
or (PTL=$medium$ and priority of $e$ is $high$ or $medium$)) then 

$\quad\quad$$P$ $\leftarrow$ $P\cup\{c\}$ 

$\quad$\textbf{End if}

\textbf{End for }

\textbf{If} (TDL\textgreater 1) \textbf{then} 

\textbf{$\quad$If} (PTL=$high$) \textbf{then }$\mathcal{A}$ $\leftarrow$$A_{h}$
\textbf{End if}

\textbf{$\quad$If} (PTL=$medium$) \textbf{then }$\mathcal{A}$ $\leftarrow$$A_{h}\cup A_{m}$
\textbf{End if}

\textbf{$\quad$For} (each $a\in\mathcal{A}$) \textbf{do} 

$\quad\quad$\textbf{If} ($a$ is not contained in any path of $P$)
\textbf{then} 

$\quad\quad\quad$$P$ $\leftarrow$ $P\cup\{a\}$ 

$\quad\quad$\textbf{End if}

$\quad$\textbf{End for}

\textbf{End if }

\caption{\label{alg:SelectRelevantTDLPaths(D,ALLTDL,TDL,PTL)}SelectRelevantTDLPaths($D$,
$ALLTDL$, TDL, PTL) \textbf{$\qquad$ Output:} $P$}
\end{algorithm}

\begin{algorithm}
$T$ $\leftarrow$ $\emptyset$

\textbf{While} (PTAB contains any elements) \textbf{do}

$\quad$$b$ $\leftarrow$ SelectBestE2EPath(PTAB, $ALLE2E$)

$\quad$$ALLE2E$ $\leftarrow$ RemoveUnnecessaryE2EPaths(PTAB, $ALLE2E$)

$\quad$PTAB $\leftarrow$ RemoveUsedTDLPaths(PTAB, $b$) 

$\quad$$T$ $\leftarrow$ $T\cup b$

\textbf{End While }

\caption{\label{alg:CreateTestCases(PTAB,ALLE2E)}CreateTestCases(PTAB, $ALLE2E$)
\textbf{$\qquad$ Output:} test set $T$ }
\end{algorithm}

\begin{algorithm}
$bestE2EPath$ $\leftarrow$ $\emptyset$

bestScore $\leftarrow$ 0 

\textbf{For} (each $x\in ALLE2E$) \textbf{do} 

$\quad$score $\leftarrow$ 0 

$\quad$\textbf{For} (each key $k$ from PTAB) \textbf{do} 

$\quad\quad$$P_{k}$ $\leftarrow$ set of all paths for key $k$
from PTAB

$\quad$$\quad$\textbf{For} (each $p\in P_{k}$) \textbf{do} 

$\quad$$\quad$$\quad$\textbf{If} ($p$ is sub-path of $x$) \textbf{then} 

$\quad$$\quad$$\quad$$\quad$score $\leftarrow$ score + 1 

$\quad$$\quad$$\quad$\textbf{End if }

$\quad$$\quad$\textbf{End for }

$\quad$$\quad$\textbf{If} (score\textgreater bestScore) \textbf{then} 

$\quad$$\quad$$\quad$bestScore $\leftarrow$ score

$\quad$$\quad$$\quad$$bestE2EPath$ $\leftarrow$ \{ $p$ \} 

$\quad$$\quad$\textbf{End if }

$\quad$\textbf{End for} 

\textbf{End for }

\caption{\label{alg:SelectBestE2EPath(PTAB,ALLE2E)}SelectBestE2EPath(PTAB,
$ALLE2E$)\textbf{$\qquad$ Output:} $bestE2EPath$ }
\end{algorithm}

\begin{algorithm}
\textbf{For} (each $x\in ALLE2E$) \textbf{do} 

$\quad$score $\leftarrow$ 0 

$\quad$\textbf{For} (each key $k$ from PTAB) \textbf{do} 

$\quad$$\quad$$P_{k}$ $\leftarrow$ set of all paths for key $k$ 

$\quad$$\quad$\textbf{For} (each $p\in P_{k}$ ) \textbf{do} 

$\quad$$\quad$$\quad$\textbf{If} ($p$ is sub-path of $x$) \textbf{then} 

$\quad$$\quad$$\quad$$\quad$score $\leftarrow$ score + 1

$\quad$$\quad$$\quad$\textbf{End if }

$\quad$$\quad$\textbf{End for }

$\quad$$\quad$\textbf{If} (score=0) \textbf{then} 

$\quad$$\quad$$\quad$$ALLE2E$ $\leftarrow$ $ALLE2E$ -- \{ $x$
\} 

$\quad$$\quad$\textbf{End if}

$\quad$\textbf{End for}

\textbf{End for} 

\caption{\label{alg:RemoveUnnecessaryE2EPaths(PTAB,ALLE2E)}RemoveUnnecessaryE2EPaths(PTAB,
$ALLE2E$)\textbf{$\qquad$ Output:} $ALLE2E$}
\end{algorithm}

\begin{algorithm}
\textbf{For} (each key $k$ from PTAB) \textbf{do}

$\quad$$P_{k}$ $\leftarrow$ set of all paths for key $k$ 

$\quad$\textbf{For} (each $p\in P_{k}$ ) \textbf{do}

$\quad$$\quad$\textbf{If} ($p$ is sub-path of $b$) \textbf{then}

$\quad$$\quad$$\quad$$P_{k}$ = $P_{k}$ -- \{ $p$ \}; 

$\quad$$\quad$$\quad$Remove $p$ from PTAB; 

$\quad$$\quad$\textbf{End if}

$\quad$\textbf{End for} 

$\quad$\textbf{If} ($P_{k}$=$\emptyset$) \textbf{then}

$\quad$$\quad$Remove key $k$ from PTAB; 

$\quad$\textbf{End if}

\textbf{End for} 

\caption{\label{alg:RemoveUsedTDLPaths(PTAB,b)}RemoveUsedTDLPaths(PTAB, $b$)
\textbf{$\qquad$ Output:} PTAB}
\end{algorithm}

We implemented the PPT algorithms in Oxygen\footnote{http://still.felk.cvut.cz/oxygen/} (formerly PCTgen) which is a model-based testing platform being developed by our research group \cite{bures2015pctgen}. For workflow testing, the PCTgen platform supports either a directed multigraph $\mathcal{G}$ with prioritization of the SUT model edges specified in Section \ref{subsec:Prioritization-of-SUT} or a simplified UML activity diagram that is converted to a weighted directed multigraph $\mathcal{G}$ before the test generation process. We used this platform to compare the algorithms in the following experiments.

\begin{figure}
\begin{centering}
\includegraphics[scale=0.7]{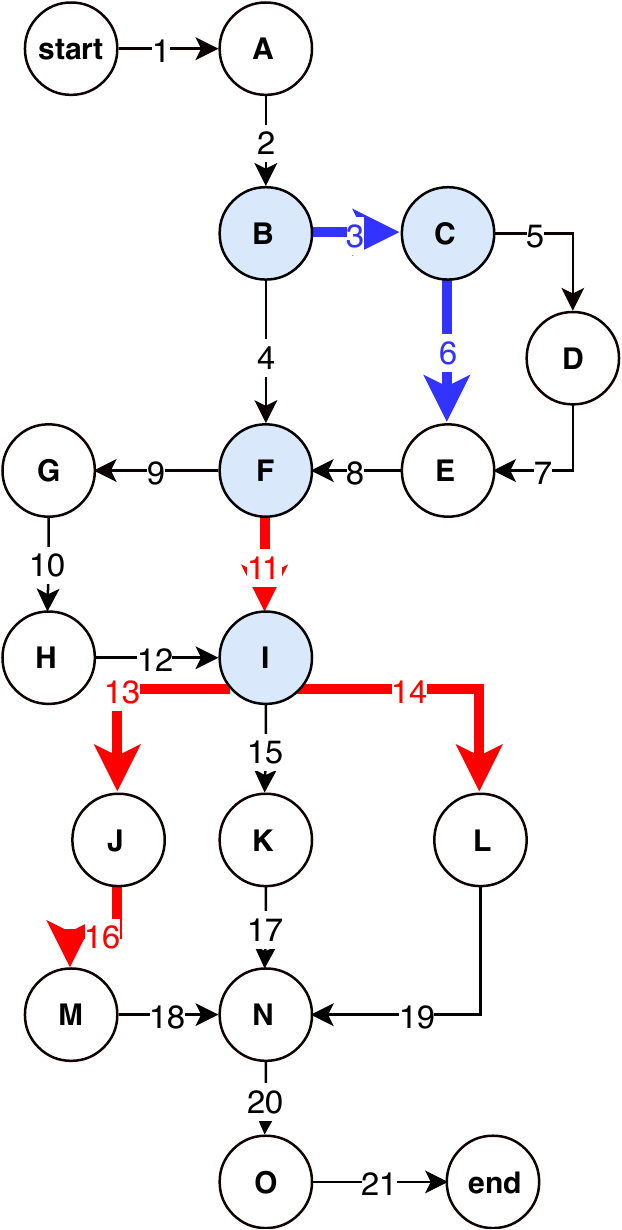}\caption{\label{fig:The-example-SUT}The running example SUT model with emphasized
priority edges}
\par\end{centering}
\end{figure}

Let us illustrate Algorithm \ref{alg:GenerateTestCases(G,TDL,PTL)} using our example from Figure \ref{fig:Coversion-of-UML}. In the example, let $A_{h}=\{$11, 13, 14, 16$\}$, $A_{m}=$\{3, 6\}, and, by default, $A_{l}=\{$1, 2, 4, 5, 7, 8, 9, 10, 12, 15, 17, 18, 19, 20, 21$\}$. In Figure \ref{fig:The-example-SUT} we depict the SUT model with $A_{h}$ emphasized in red and $A_{m}$ emphasized in blue.

During the execution of Algorithm \ref{alg:GenerateTestCases(G,TDL,PTL)},
the set $ALLTDL$ (line 6) is equivalent to $A_{R}$ for $TDL=1$
and to a set of paths $\{$1-2, 2-3, 2-4, 3-5, 3-6, 4-9, 4-11, 5-7, 6-8, 7-8, 8-9, 8-11, 9-10, 10-12, 11-13, 11-14, 11-15, 12-13, 12-14, 12-15, 13-16, 14-19, 15-17, 16-18, 17-20, 18-20, 19-20, 20-21$\}$ for $TDL=2$.

After subsequent selection of relevant paths from $ALLTDL$ (line 8), the set $P$ is equivalent to \{{11, 13, 14, 16}\}  for $TDL=1$ and for $PTL=high$. For $TDL=1$ and $PTL=medium$, $P$ is equivalent to \{3, 6, 11, 13, 14, 16\}. For $TDL=2$ and $PTL=high$, $P$ is equivalent to \{11-13, 11-14, 11-15, 13-16, 14-19, 16-18\}. Finally, for $TDL=2$ and $PTL=medium$, $P$ is equivalent to \{3-5, 3-6, 6-8, 11-13, 11-14, 11-15, 13-16, 14-19, 16-18\}.

For $TDL=1$ and $PTL=high$, Algorithm \ref{alg:GenerateTestCases(G,TDL,PTL)}
produces a test set $T_{R1H}=\{t_{1H1},t_{1H2}\}$, where $t_{1H1}=\{$start, 1, A, 2, B, 4, F, 11, I, 13, J, 16, M, 18, N, 20, O, 21, end$\}$ and $t_{1H2}=\{$start, 1, A, 2, B, 4, F, 11, I, 14, L, 19, N, 20, O, 21, end $\}$. For $TDL=1$ and for $PTL=medium$, the algorithm produces $T_{R1M}=\{t_{1M1},t_{1M2}\}$,
where $t_{1M1}=\{$start, 1, A, 2, B, 3, C, 6, E, 8, F, 11, I, 13, J, 16, M, 18, N, 20, O, 21, end $\}$ and $t_{1M2}=\{$start, 1, A, 2, B, 4, F, 11, I, 14, L, 19, N, 20, O, 21, end$\}$. Note that the test case $t_{1M1}$ is longer than $t_{1H1}$, as nodes {C} and {E} must be visited due to $PTL=medium$.

For $TDL=2$ and $PTL=high$, the algorithm produces $T_{R2H}=\{t_{2H1},t_{2H2},t_{2H3}\}$, where $t_{2H1}=\{$start, 1, A, 2, B, 4, F, 11, I, 13, J, 16, M, 18, N, 20, O, 21, end$\}$, $t_{2H2}=\{$start, 1, A, 2, B, 4, F, 11, I, 14, L, 19, N, 20, O, 21, end$\}$, and $t_{2H3}=\{$start, 1, A, 2, B, 4, F, 11, I, 15, K, 17, N, 20, O, 21, end$\}$. Compared to $T_{R1H}$, $\mid T_{R2H}\mid=3$, as in $T_{R2H}$ edge {15} must be visited because $TDL=2$ (path \{15-K-17-N\} is sideway of the high priority part of the process modeled by the edges {11}, {13}, {14} and {16}).

Finally, for $TDL=2$ and $PTL=medium$, the algorithm produces $T_{R2M}=\{t_{2M1},t_{2M2},t_{2M3}\}$, where $t_{2M1}=\{$start, 1, A, 2, B, 3, C, 6, E, 8, F, 11, I, 13, J, 16, M, 18, N, 20, O, 21, end$\}$, $t_{2M2}=\{$start, 1, A, 2, B, 3, C, 5, D, 7, E, 8, F, 11, I, 14, L, 19, N, 20, O, 21, end$\}$,
and $t_{2M3}=\{$start, 1, A, 2, B, 4, F, 11, I, 15, K, 17, N, 20, O, 21, end$\}$.
Compared to $T_{R1M}$ and $T_{R2H}$, in $T_{R2M}$, the node {D}
must be visited because $TDL=2$ (path {5-D-7} is sideway of
the medium priority part of the process modelled by edges {3} and {6}).

\section{\label{sec:Experiments}Experimental Evaluation}

In the experimental evaluation of the Prioritized Process Test, we conducted a comparison with five alternative algorithms. In the experiments, we compared:

\begin{enumerate} 

\item The Process Cycle Test (PCT) without a prioritization mechanism \cite{koomen2013tmap} (in the comparison, the results of this algorithm are used only as a baseline to provide an idea of the size of the test sets without prioritization). PCT uses $\mathcal{G}$ as an SUT model. 
\item A simulation of a naive method for prioritization of the test cases using $\mathcal{G}$ as an SUT model. In this method, $T$ is a set of PCT test cases that has been reduced by the removal of all test cases that do not contain: 
\begin{enumerate}
\item any edge with priority $high$ for $PTL=high$ (further denoted as
DCT(h)),
\item any edge with priority $high$ or $medium$ for $PTL=medium$ (further
denoted as DCT(m)).
\end{enumerate}
\item Test cases produced by the Prioritized Process Test (PPT) proposed in Section \ref{sec:Prioritized-Process-Test} using $\mathcal{G}$ as an SUT model (further denoted as PPT(h) for $PTL=high$ and PPT(m) for $PTL=medium$).
\item The Brute Force algorithm using $G$ and a set of test requirements $R$ as an SUT model, published by Li \textit{et al.} \cite{li2012better} (further denoted as BF). 
\item The Set-Covering Based Solution using $G$ and $R$ \cite{li2012better}
(further denoted as SC).
\item The Matching-Based Prefix Graph Solution, also using $G$ and $R$
as an SUT model \cite{li2012better} (further denoted as PG). 
\end{enumerate}

All algorithms PCT, DCT, PPT, BF, SC and PG were implemented in the PCTgen platform. Regarding the BF algorithm, we used our own implementation based on the pseudocode published in \cite{li2012better}. Here, we exactly implemented the pseudocode without any changes or optimizations of the algorithm. As BF composes the test case as a sequence of $G$ nodes $d_{1},d_{2},..,d_{n}$, we added a conversion mechanism to transform the produced test cases to sequences of $G$ edges $a_{1},a{}_{2},..,a_{n-1}$, to ensure comparability of the BF results with those of PCT, DCT and PPT. Regarding the SC and PG algorithms, we used open-source code published by Offut et al. \cite{offuttools} and we integrated this code with the Oxygen platform.

Several alternative approaches mentioned in this paper are related to the discussed problem. However, comparability with the presented PPT algorithm is affected by the following issues. Regarding the work by Dwarakanath and Jankiti \cite{dwarakanath2014minimum}, they focus on prime-path coverage, which is out of the scope of the paper. The approach by Gökçe et al. \cite{gokcce2006coverage} (also presented in Belli et al.\cite{belli2007coverage}) is aimed at minimization
of the test set for complete coverage, which is also out of the scope of the paper. Moreover, prioritization by individual parts of the SUT as discussed in this paper is not the subject of the study. Panthi and Mohapatra \cite{panthi2015generating}aim at optimization of all possible feasible test sequences in a control flow graph generated for a state machine diagram modelling the SUT. Despite the similarity with the scope of this paper, as this paper focuses on of all possible feasible test sequences, its scope differs from the test coverage viewpoint.

\subsection{Experiment Method}

As the testing data in the experiments, we used a set of created graphs $\mathcal{G}$ representing an SUT model. The graphs were manually created to correspond to workflows of three real software systems:
the medical information system Pluto, the customer relationship management system Global and the issue tracking system MantisBT. The graphs capture processes and workflows on two principal levels in these three systems:
(1) high-level business workflow and (2) low-level technical workflow on the code level. In all cases, either running development version, the source code and design documentation (for Pluto and Global) or the source code (for MantisBT) of these systems are available to our research team members; hence, the relevance of the created graphs to a real-world software development problem was ensured.

Because BF,SC and PG use $G$ as an SUT model, to ensure objective comparability of PCT, DCT, PPT, BF, SC and PG, the graphs do not contain parallel edges; therefore, the problem instances were not directed multigraphs, but were rather directed graphs only.

To compare PCT, DCT and PPT with BF, SC and PG, each of the graphs $\mathcal{G}$ was converted to a corresponding graph $G$ and a set of test requirements $R$ (defined as sequences of $G$ nodes that must be present in the test cases) by the following process:
\begin{enumerate}
\item $G$ is created from $\mathcal{G}$ by simply neglecting its edge weights.
\item For $TDL=1$, $R$ is created from $\mathcal{G}$ by two alternative methods, atomic conversion and sequence conversion, as specified in Table \ref{tab:Test_requirements_conversion_specification}. Both alternative methods are evaluated in the experiments. 
\item For $TDL=2$, which is practically equivalent to the \textit{Edge-pair Coverage} criterion, the process is as follows. A set $A_{pair}$ contains all possible pairs of adjacent edges of $\mathcal{G}$. Then, $R$ is a set of all paths $(d_{i},d_{i+1},d_{i+2})$, such that $a_{i}=(d_{i,}d_{i+1})$, $a_{i+1}=(d_{i+1,}d_{i+2})$ for each $(a_{i},a{}_{i+1})\in A_{pair}$.
\end{enumerate}

\begin{table}
\begin{centering}
\caption{\label{tab:Test_requirements_conversion_specification}Method of creation
of the test requirements $R$ from the graph $\mathcal{G}$}
\par\end{centering}
\centering{}%
\begin{tabular}{|>{\centering}p{1.8cm}|>{\centering}p{3cm}|>{\centering}p{3cm}|}
\hline 
Method of creation of $R$ from $\mathcal{G}$  & $PTL=high$ & $PTL=medium$\tabularnewline
\hline 
\hline 
Atomic conversion & $R$ = set of all $G$ adjacent node pairs $a=(d_{i},d_{i+1})$ for
each $a\in A_{h}$ & $R$ = set of all $G$ adjacent node pairs $a=(d_{i},d_{i+1})$ for
each $a\in A_{h}\cup A_{m}$ \tabularnewline
\hline 
Sequence conversion & $R$ = set of paths in $\mathfrak{\mathcal{G}}$, each $a\in p\in R$
has priority $high$ & $R$ = set of paths in $\mathfrak{\mathcal{G}}$, each $a\in p\in R$
has priority $high$ or $medium$\tabularnewline
\hline 
\end{tabular}
\end{table}

Illustrating the conversion process using our running example, for atomic conversion and $PTL=high$, $R$ is a set of adjacent node pairs for edges \{11, 13, 14, 16\}. For $PTL=medium$, $R$ is a set of adjacent node pairs for edges \{3, 6, 11, 13, 14, 16\}. For sequence conversion and $PTL=high$, $R$ is a set of paths \{11-13-16, 11-14\} and for $PTL=medium$, $R$ is a set of paths \{3-6, 11-13-16, 11-14\}.

Properties of the SUT models used in the experiments are presented in Table \ref{tab:Selected-SUT-models}. Value $loops$ denote the number of loops and value $deg$ denotes the average node degree (the sum of the average outgoing node degree and the average incoming node degree) in a particular graph. Values$\mid R\mid_{PTL=high}$ and $\mid R\mid_{PTL=medium}$ in the table \ref{tab:Selected-SUT-models} are given for the sequence conversion of the test requirements (see Table \ref{tab:Test_requirements_conversion_specification}). For
the atomic conversion of test requirements $\mid R\mid_{PTL=high}=\mid A_{h}\mid$ and $\mid R\mid_{PTL=medium}=\mid A_{h}\mid+\mid A_{m}\mid.$

\begin{table*}
\scriptsize

\caption{\label{tab:Selected-SUT-models}SUT models used in experiments}

\centering{}%
\begin{tabular}{|c|c|c|c|c|c|c|c|c|c|}
\hline 
SUT model ID & $\mid D\mid$ & $\mid A\mid$ & $\mid A_{h}\mid$ & $\mid A_{m}\mid$ & $\mid A_{l}\mid$ & $loops$ & $\mid R\mid_{PTL=high}$ & $\mid R\mid_{PTL=medium}$ & $deg$\tabularnewline
\hline 
\hline 
1 & 11 & 19 & 4 & 2 & 13 & 5 & 2 & 4 & 3.45\tabularnewline
\hline 
2 & 13 & 19 & 4 & 2 & 13 & 5 & 3 & 5 & 2.92\tabularnewline
\hline 
3 & 24 & 43 & 9 & 8 & 26 & 10 & 7 & 12 & 3.58\tabularnewline
\hline 
4 & 15 & 24 & 8 & 5 & 11 & 7 & 4 & 7 & 3.20\tabularnewline
\hline 
5 & 14 & 22 & 3 & 3 & 16 & 7 & 2 & 5 & 3.14\tabularnewline
\hline 
6 & 9 & 14 & 5 & 1 & 8 & 4 & 3 & 4 & 3.11\tabularnewline
\hline 
7 & 13 & 21 & 4 & 3 & 14 & 7 & 3 & 6 & 3.23\tabularnewline
\hline 
8 & 15 & 23 & 6 & 6 & 11 & 6 & 6 & 10 & 3.07\tabularnewline
\hline 
9 & 13 & 19 & 5 & 2 & 12 & 5 & 3 & 4 & 2.92\tabularnewline
\hline 
10 & 19 & 32 & 6 & 4 & 22 & 7 & 6 & 7 & 3.37\tabularnewline
\hline 
11 & 15 & 25 & 3 & 4 & 18 & 8 & 2 & 3 & 3.33\tabularnewline
\hline 
12 & 16 & 26 & 7 & 3 & 16 & 9 & 7 & 8 & 3.25\tabularnewline
\hline 
13 & 12 & 19 & 6 & 2 & 11 & 5 & 3 & 5 & 3.17\tabularnewline
\hline 
14 & 14 & 22 & 6 & 3 & 13 & 8 & 4 & 5 & 3.14\tabularnewline
\hline 
15 & 16 & 19 & 5 & 5 & 9 & 0 & 4 & 5 & 2.38\tabularnewline
\hline 
16 & 6 & 10 & 3 & 3 & 4 & 1 & 3 & 2 & 3.33\tabularnewline
\hline 
17 & 11 & 16 & 2 & 3 & 11 & 0 & 2 & 5 & 2.91\tabularnewline
\hline 
18 & 13 & 20 & 3 & 4 & 13 & 0 & 2 & 6 & 3.08\tabularnewline
\hline 
19 & 8 & 10 & 1 & 3 & 6 & 2 & 1 & 4 & 2.50\tabularnewline
\hline 
20 & 9 & 11 & 2 & 3 & 6 & 0 & 1 & 2 & 2.44\tabularnewline
\hline 
21 & 10 & 15 & 3 & 3 & 9 & 0 & 2 & 3 & 3.00\tabularnewline
\hline 
22 & 7 & 9 & 1 & 4 & 4 & 0 & 1 & 4 & 2.57\tabularnewline
\hline 
23 & 8 & 12 & 2 & 3 & 7 & 0 & 2 & 3 & 3.00\tabularnewline
\hline 
24 & 10 & 12 & 3 & 2 & 7 & 0 & 3 & 3 & 2.40\tabularnewline
\hline 
25 & 8 & 12 & 3 & 2 & 7 & 3 & 2 & 2 & 3.00\tabularnewline
\hline 
26 & 8 & 11 & 3 & 3 & 5 & 3 & 3 & 5 & 2.75\tabularnewline
\hline 
27 & 7 & 12 & 3 & 2 & 7 & 5 & 3 & 3 & 3.43\tabularnewline
\hline 
28 & 8 & 11 & 2 & 4 & 5 & 2 & 2 & 4 & 2.75\tabularnewline
\hline 
29 & 7 & 11 & 4 & 2 & 5 & 0 & 2 & 2 & 3.14\tabularnewline
\hline 
30 & 10 & 15 & 3 & 4 & 8 & 1 & 2 & 3 & 3.00\tabularnewline
\hline 
31 & 23 & 32 & 7 & 9 & 16 & 3 & 6 & 13 & 2.78\tabularnewline
\hline 
32 & 26 & 40 & 8 & 4 & 28 & 4 & 8 & 9 & 3.08\tabularnewline
\hline 
33 & 35 & 48 & 5 & 9 & 34 & 4 & 5 & 11 & 2.74\tabularnewline
\hline 
34 & 45 & 61 & 10 & 9 & 42 & 5 & 8 & 15 & 2.71\tabularnewline
\hline 
35 & 21 & 27 & 12 & 6 & 9 & 0 & 2 & 3 & 2.57\tabularnewline
\hline 
36 & 19 & 24 & 7 & 4 & 13 & 1 & 3 & 4 & 2.53\tabularnewline
\hline 
37 & 24 & 29 & 8 & 9 & 12 & 2 & 5 & 7 & 2.42\tabularnewline
\hline 
38 & 25 & 35 & 8 & 7 & 20 & 0 & 5 & 4 & 2.80\tabularnewline
\hline 
39 & 26 & 38 & 10 & 3 & 25 & 2 & 3 & 4 & 2.92\tabularnewline
\hline 
40 & 27 & 37 & 8 & 7 & 22 & 3 & 8 & 12 & 2.74\tabularnewline
\hline 
41 & 14 & 20 & 5 & 5 & 10 & 1 & 5 & 6 & 2.86\tabularnewline
\hline 
42 & 21 & 26 & 3 & 3 & 20 & 0 & 3 & 5 & 2.48\tabularnewline
\hline 
43 & 20 & 30 & 7 & 4 & 19 & 4 & 4 & 5 & 3.00\tabularnewline
\hline 
44 & 28 & 46 & 13 & 10 & 23 & 5 & 11 & 19 & 3.29\tabularnewline
\hline 
45 & 21 & 28 & 10 & 6 & 12 & 0 & 7 & 12 & 2.67\tabularnewline
\hline 
46 & 19 & 31 & 9 & 9 & 13 & 6 & 8 & 19 & 3.26\tabularnewline
\hline 
47 & 25 & 39 & 9 & 11 & 19 & 8 & 8 & 13 & 3.12\tabularnewline
\hline 
48 & 52 & 79 & 7 & 9 & 63 & 3 & 5 & 10 & 3.04\tabularnewline
\hline 
49 & 47 & 68 & 12 & 8 & 48 & 3 & 5 & 10 & 2.89\tabularnewline
\hline 
50 & 46 & 65 & 9 & 11 & 45 & 0 & 6 & 10 & 2.83\tabularnewline
\hline 
51 & 61 & 97 & 21 & 10 & 66 & 3 & 12 & 17 & 3.18\tabularnewline
\hline 
52 & 51 & 71 & 16 & 8 & 47 & 0 & 10 & 13 & 2.78\tabularnewline
\hline 
53 & 27 & 40 & 11 & 3 & 26 & 2 & 8 & 10 & 2.96\tabularnewline
\hline 
54 & 21 & 22 & 7 & 4 & 11 & 0 & 4 & 5 & 2.10\tabularnewline
\hline 
55 & 29 & 35 & 9 & 8 & 18 & 0 & 4 & 8 & 2.41\tabularnewline
\hline 
56 & 34 & 50 & 10 & 8 & 32 & 0 & 6 & 9 & 2.94\tabularnewline
\hline 
57 & 35 & 50 & 8 & 4 & 38 & 0 & 8 & 11 & 2.86\tabularnewline
\hline 
58 & 37 & 55 & 16 & 5 & 34 & 2 & 9 & 13 & 2.97\tabularnewline
\hline 
59 & 35 & 48 & 12 & 8 & 28 & 1 & 10 & 11 & 2.74\tabularnewline
\hline 
\end{tabular}
\end{table*}

To compare the test sets produced by the individual strategies, we
use two sets of metrics: the test set metrics and efficiency metrics.
The \textbf{test set metrics} are based on the properties of the test
set $T$ produced by a particular strategy:
\begin{itemize}
\item $\mid T\mid$ - number of tests in a generated test set
\item $\alpha$ - total number of edges in all test cases of a test set
$T$
\item $\alpha_{h}$ - total number of edges of priority $high$ in all test
cases of a test set $T$
\item $\alpha_{m}$ - total number of edges of priority $high$ and $medium$
in all test cases of a test set $T$
\item $\beta$ - total number of unique edges in all test cases of a test
set $T$
\item $\beta_{h}$ - total number of unique edges of priority $high$ in
all test cases of a test set $T$. For correctly generated test cases
and $PTL=high$, $\beta_{h}=\mid A_{h}\mid$. This metric was used
to verify the consistency of the test cases.
\item $\beta_{m}$ - total number of unique edges of priority $high$ or
$medium$ in all test cases of a test set $T$. For correctly generated
test cases and $PTL=medium$, $\beta_{m}=\mid A_{h}\mid+\mid A_{m}\mid$.
This metric was used to verify the consistency of the test cases.
\item $\delta$ - total number of nodes in all test cases of a test set
$T$
\item $\varepsilon$- total number of unique nodes in all test cases of
a test set $T$
\end{itemize}
\textbf{The efficiency metrics} are calculated from the values of
the test set metrics and parameters of the SUT model $G$. This set
of metrics reflects more on the efficiency of test cases of the test
set $T$.
\begin{itemize}
\item $ac=\frac{\beta}{\mid A\mid}.100\%$ - ratio of unique edges contained
in a test set $T$ (in percentage). For PCT, $ac=100\%$ by the principle
of the algorithm. A lower value of $ac$ indicates more optimal test
cases. Fewer unique edges (which would imply extra testing costs)
are present in the test cases while the required test coverage criteria
are maintained. 
\item $\lambda_{h}=\frac{\alpha_{h}}{\alpha}.100\%$ - ratio of edges of
priority $high$ and all edges contained in a test set $T$ (in percentage).
A higher value of $\lambda_{h}$ indicates more optimal test cases:
fewer edges that do not have priority $high$ (and thus are not necessary
to test) are present in the test cases while the required coverage
criteria are maintained.
\item $\varLambda_{h}=\frac{\beta_{h}}{\alpha}.100\%$ - ratio of unique
edges of priority $high$ and all edges contained in a test set $T$
(in percentage). A higher value of $\varLambda_{h}$ indicates more
optimal test cases. Fewer unique edges that do not have priority $high$
(and thus are not necessary to test) are present in the test cases
while the required coverage criteria are maintained.
\item $\lambda_{m}=\frac{\alpha_{m}}{\alpha}.100\%$ - ratio of edges of
priority $high$ or $medium$ and all edges contained in a test set
$T$ (in percentage). A higher value of $\lambda_{m}$ indicates more
optimal test cases. Fewer edges that do not have priority $high$
or $medium$ (and thus are not necessary to test) are present in the
test cases while the required coverage criteria are maintained.
\item $\varLambda_{m}=\frac{\beta_{m}}{\alpha}.100\%$ - ratio of unique
edges of priority $high$ and $medium$ and all edges contained in
a test set $T$ (in percentage). A higher value of $\varLambda_{m}$
indicates more optimal test cases. Fewer unique edges that do not
have priority $high$ or $medium$ (and thus are not necessary to
test) are present in the test cases while the required coverage criteria
are maintained.
\end{itemize}
In the experiments, we ran the PCT, DCT(h), DCT(m), PPT(h) and PPT(m)
algorithms for $TDL=1$ and $TDL=2$, which we consider the test coverage
being used in the majority of testing assignments for noncritical
software systems \cite{van2013tmap}. For $TDL=1$, the BF, SC and
PG algorithms were executed for all four combinations of $PTL$, and
the method of creation for a set of test requirements $R$, as specified
in Table \ref{tab:Test_requirements_conversion_specification}. For
$TDL=2$, the BF, SC and PG algorithms were executed with a set of
test requirements that contains all pairs of adjacent edges (equivalent
to o $TDL=2$). For $TDL=1$ and $TDL=2$, PCT acts only as a baseline
for comparison, as this algorithm does not reflect either prioritization
of $\mathcal{G}$ edges, nor a set of test requirements $R$. For
$TDL=2$, BF, SC and PG also take the role of a comparison baseline,
as the set of test requirements $R$ is used to model the requirements
of \textit{Edge-pair Coverage}.

Test cases produced by all algorithms were automatically verified
for their consistency and satisfaction of test requirements by the
following checks:

\begin{enumerate}
\item A test case $t\in T$ is a path in $\mathcal{G}$ (or $G$); which starts in $d_{s}$ and ends in any node of $D_{e}$ (applies to all algorithms),
\item  for $PTL=high$, all edges of $A_{h}$ are present in the test cases, for $PTL=medium$, all edges of $A_{h}\cup A_{m}$ are present in the test cases (applies to the PCT, DCT and PPT algorithms), 
\item all test requirements of $R$ are present in the test cases (applies to the BF, SC and PG algorithms), and 
\item for $PTL=high$, $\beta_{h}=\mid A_{h}\mid$and for $PTL=medium$, $\beta_{h}=\mid A_{h}\mid$ and $\beta_{m}=\mid A_{h}\mid+\mid A_{m}\mid$ (applies to all algorithms).
\end{enumerate}

\subsection{Experimental Results}

During the experiments, all algorithms were run on the same hardware
and software configuration. This configuration was an Intel i5 2.40GHz
CPU, 8GB RAM, Ubuntu 16.04.3 operating system, OpenJDK Runtime Environment
version 1.8.0.

In Tables \ref{tab:results_TDL=00003D1_PTL=00003Dhigh}, \ref{tab:results_TDL=00003D1_PTL=00003Dmedium},
\ref{tab:results_TDL=00003D2_PTL=00003Dhigh} and \ref{tab:results_TDL=00003D2_PTL=00003Dmedium},
$time$ denotes the execution time of the algorithm in milliseconds.
For the BF algorithm, the time for generation of $R$ from $\mathfrak{\mathcal{G}}$
edge priorities was not included in the measured execution time to
make the comparison more objective, as $R$ generation represents
a conversion step only and precedes the generation of the test cases.
For the DCT test cases are considered as inputs to the test case reduction
process, and the measured time covers only this reduction. Thus, the
total execution time of DCT shall also include the time needed for
the generation of PCT test cases.

Table \ref{tab:results_TDL=00003D1_PTL=00003Dhigh} summarizes the
results for PCT, DCT(h), PPT(h),\textbf{ }BF, SC and PG with $R$
created by the atomic and sequence conversions for $TDL=1$ and $PTL=high$.
In Table \ref{tab:results_TDL=00003D1_PTL=00003Dhigh}, the averaged
results of all 59 graphs presented in Table \ref{tab:Selected-SUT-models}
are provided. In the following tables, \textbf{BF} \textbf{a} denotes
the BF algorithm with $R$ created by the atomic conversion, and \textbf{BF}
\textbf{s} denotes the BF algorithm with $R$ created by the sequence
conversion. The same notation is used for the SC and PG algorithms
(\textbf{SC} \textbf{a},\textbf{ SC} \textbf{s}, \textbf{PG} \textbf{a}
and\textbf{ PG} \textbf{s}).

The comparison of the individual algorithms is presented in Figure
\ref{fig:Comparison-of-algorithms-TDL=00003D1, PTL=00003Dhigh}. In
Tables \ref{tab:results_TDL=00003D1_PTL=00003Dhigh}, \ref{tab:results_TDL=00003D1_PTL=00003Dmedium},
\ref{tab:results_TDL=00003D2_PTL=00003Dhigh} and \ref{tab:results_TDL=00003D2_PTL=00003Dmedium},
the best scores for individual metrics are emphasized in bold. Detailed
data for all graphs are available in Appendices A-D.

\begin{table*}
\scriptsize

\caption{\label{tab:results_TDL=00003D1_PTL=00003Dhigh}Experiment results
for TDL=1, PTL=high. Detailed data for all graphs are available in
Appendix A.}

\begin{singlespace}
\centering{}%
\begin{tabular}{|c|c|c|c|c|c|c|c|c|c|}
\hline 
Metric / $\mathcal{G}$  & \textbf{PCT} & \textbf{DCT(h)} & \textbf{PPT(h)} & \textbf{BF} \textbf{a} & \textbf{BF} \textbf{s} & \textbf{SC} \textbf{a} & \textbf{SC} \textbf{s} & \textbf{PG} \textbf{a} & \textbf{PG} \textbf{s}\tabularnewline
\hline 
\hline 
$\mid T\mid$ & 9.86 & 8.10 & \textbf{2.56} & 4.85 & 4.19 & 4.69 & 3.88 & 3.93 & 3.80\tabularnewline
\hline 
$\alpha$ & 85.97 & 78.00 & \textbf{19.90} & 32.93 & 29.20 & 32.76 & 27.95 & 27.92 & 27.42\tabularnewline
\hline 
$\alpha_{h}$ & 21.51 & 21.51 & \textbf{7.86} & 11.31 & 10.42 & 11.59 & 10.15 & 9.95 & 10.05\tabularnewline
\hline 
$\alpha_{m}$ & 34.83 & 33.08 & \textbf{9.95} & 14.92 & 13.64 & 15.03 & 13.25 & 12.95 & 12.98\tabularnewline
\hline 
$\beta$ & 30.59 & 27.90 & \textbf{15.93} & 18.39 & 17.53 & 17.71 & 16.98 & 17.03 & 16.86\tabularnewline
\hline 
$\beta_{h}$ & \textbf{6.71} & \textbf{6.71} & \textbf{6.71} & \textbf{6.71} & \textbf{6.71} & \textbf{6.71} & \textbf{6.71} & \textbf{6.71} & \textbf{6.71}\tabularnewline
\hline 
$\beta_{m}$ & 11.78 & 11.12 & \textbf{8.41} & 8.95 & 8.81 & 8.76 & 8.63 & 8.71 & 8.56\tabularnewline
\hline 
$\delta$ & 76.10 & 69.90 & \textbf{17.34} & 28.08 & 25.02 & 28.07 & 24.07 & 23.98 & 23.63\tabularnewline
\hline 
$\varepsilon$ & 28.68 & 26.20 & \textbf{14.39} & 16.76 & 15.93 & 16.10 & 15.42 & 15.42 & 15.31\tabularnewline
\hline 
$ac$  & 100\% & 90.26\% & \textbf{53.86\%} & 61.51\% & 58.81\% & 59.04\% & 56.80\% & 57.10\% & 56.45\%\tabularnewline
\hline 
$\lambda_{h}$ & 24.28\% & 27.62\% & \textbf{40.73\%} & 34.82\% & 37.19\% & 36.72\% & 38.62\% & 38.18\% & 38.83\%\tabularnewline
\hline 
$\lambda_{m}$ & 42.52\% & 45.60\% & \textbf{52.50\%} & 47.33\% & 49.42\% & 48.52\% & 50.55\% & 50.47\% & 50.63\%\tabularnewline
\hline 
$\varLambda_{h}$ & 9.90\% & 11.91\% & \textbf{36.74\%} & 23.23\% & 27.06\% & 24.18\% & 28.98\% & 28.34\% & 29.21\%\tabularnewline
\hline 
$\varLambda_{m}$ & 18.10\% & 20.97\% & \textbf{46.62\%} & 31.68\% & 35.98\% & 31.87\% & 37.48\% & 37.11\% & 37.71\%\tabularnewline
\hline 
$time$ & 13.29 & 8.69 & 35.46 & 1.39 & \textbf{1.35} & 2.25 & 1.30 & 49.16 & 53.13\tabularnewline
\hline 
\end{tabular}
\end{singlespace}
\end{table*}

\begin{figure*}
\begin{centering}
\includegraphics[width=15cm]{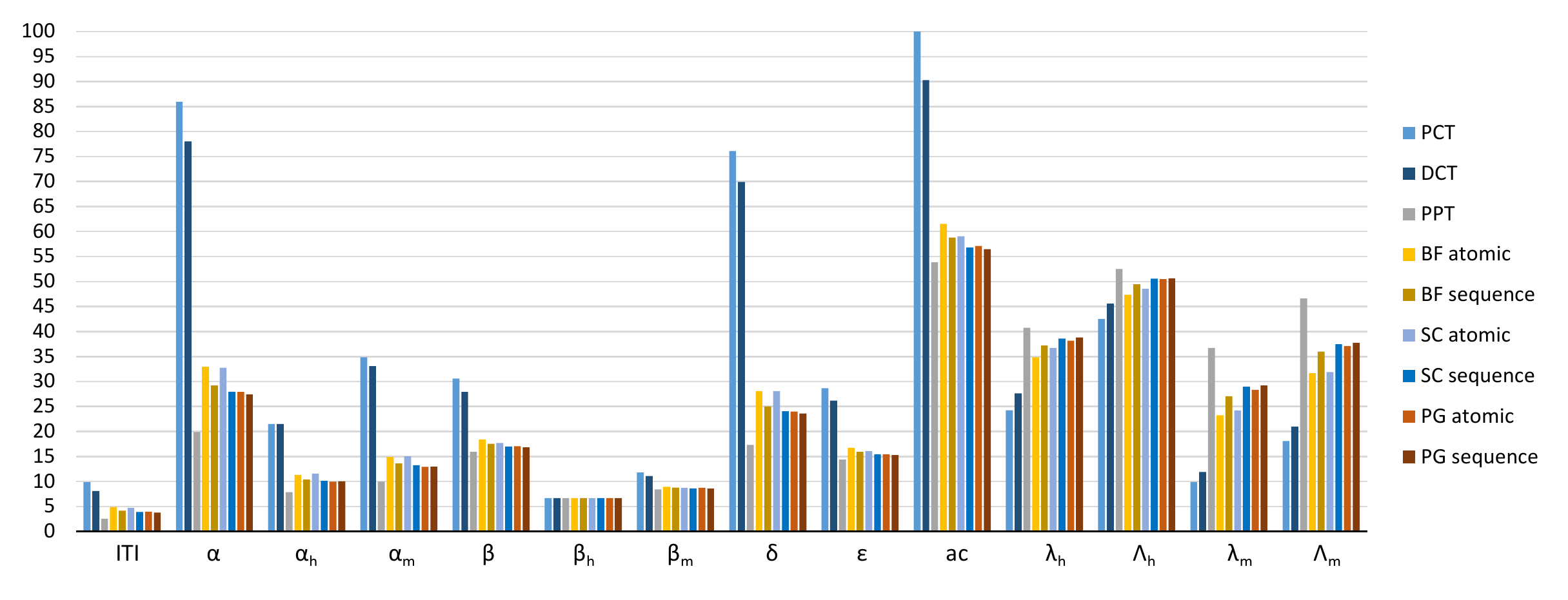}
\par\end{centering}
\caption{Comparison of algorithms for TDL=1 and PTL=high \label{fig:Comparison-of-algorithms-TDL=00003D1, PTL=00003Dhigh}}
\end{figure*}

Further, Table \ref{tab:results_TDL=00003D1_PTL=00003Dmedium} summarizes
the results of DCT(m), PPT(m), BF, SC and PG with $R$ created by
the atomic and sequence conversions for $TDL=1$ and $PTL=medium$.
In Table \ref{tab:results_TDL=00003D1_PTL=00003Dmedium}, the averaged
results for all 59 graphs presented in Table \ref{tab:Selected-SUT-models}
are provided.

\begin{figure*}
\begin{centering}
\includegraphics[width=15cm]{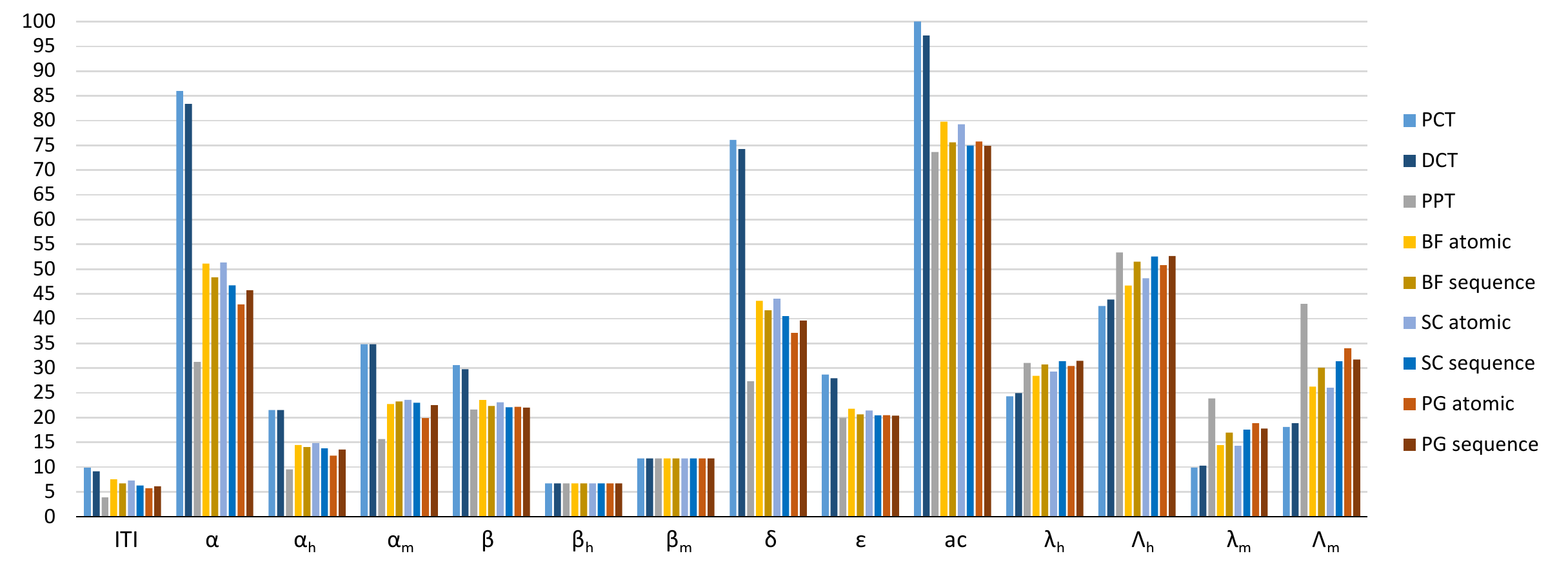}
\par\end{centering}
\caption{\label{fig:Comparison-of-algorithms for TDL=00003D1 and PTL=00003Dmedium}Comparison
of algorithms for TDL=1 and PTL=medium}
\end{figure*}

The results of the DCT(m), PPT(m), BF, SC and PG algorithms in these
test coverage criteria can be compared with PCT with $TDL=1$, acting
as a baseline when no prioritization is reflected in the test set.
A comparison of values averaged for all 59 graphs is presented in
Figure \ref{fig:Comparison-of-algorithms for TDL=00003D1 and PTL=00003Dmedium}.

\begin{table*}
\scriptsize
\centering{}\caption{\label{tab:results_TDL=00003D1_PTL=00003Dmedium}Experiment results
for TDL=1, PTL=medium. Detailed data for all graphs are available
in Appendix B.}
\begin{tabular}{|c|c|c|c|c|c|c|c|c|c|}
\hline 
Metric / $\mathcal{G}$  & \textbf{PCT} & \textbf{DCT(m)} & \textbf{PPT(m)} & \textbf{BF} \textbf{a} & \textbf{BF} \textbf{s} & \textbf{SC} \textbf{a} & \textbf{SC} \textbf{s} & \textbf{PG} \textbf{a} & \textbf{PG} \textbf{s}\tabularnewline
\hline 
\hline 
$\mid T\mid$ & 9.86 & 9.14 & \textbf{3.92} & 7.51 & 6.69 & 7.32 & 6.27 & 5.76 & 6.12\tabularnewline
\hline 
$\alpha$ & 85.97 & 83.36 & \textbf{31.27} & 51.10 & 48.36 & 51.36 & 46.75 & 42.86 & 45.75\tabularnewline
\hline 
$\alpha_{h}$ & 21.51 & 21.51 & \textbf{9.56} & 14.47 & 14.02 & 14.88 & 13.76 & 12.29 & 13.54\tabularnewline
\hline 
$\alpha_{m}$ & 34.83 & 34.83 & \textbf{15.66} & 22.73 & 23.27 & 23.58 & 23.00 & 19.90 & 22.51\tabularnewline
\hline 
$\beta$ & 30.59 & 29.76 & \textbf{21.63} & 23.54 & 22.34 & 23.10 & 22.05 & 22.20 & 22.03\tabularnewline
\hline 
$\beta_{h}$ & \textbf{6.71} & \textbf{6.71} & \textbf{6.71} & \textbf{6.71} & \textbf{6.71} & \textbf{6.71} & \textbf{6.71} & \textbf{6.71} & \textbf{6.71}\tabularnewline
\hline 
$\beta_{m}$ & \textbf{11.78} & \textbf{11.78} & \textbf{11.78} & \textbf{11.78} & \textbf{11.78} & \textbf{11.78} & \textbf{11.78} & \textbf{11.78} & \textbf{11.78}\tabularnewline
\hline 
$\delta$ & 76.10 & 74.22 & \textbf{27.36} & 43.59 & 41.66 & 44.03 & 40.47 & 37.10 & 39.63\tabularnewline
\hline 
$\varepsilon$ & 28.68 & 27.93 & \textbf{19.95} & 21.81 & 20.68 & 21.41 & 20.42 & 20.53 & 20.41\tabularnewline
\hline 
$ac$  & 100\% & 97.13\% & \textbf{73.64\%} & 79.77\% & 75.58\% & 79.22\% & 74.95\% & 75.78\% & 74.91\%\tabularnewline
\hline 
$\lambda_{h}$ & 24.28\% & 24.97\% & 31.06\% & 28.41\% & 30.69\% & 29.32\% & 31.41\% & 30.44\% & \textbf{31.49\%}\tabularnewline
\hline 
$\lambda_{m}$ & 42.52\% & 43.86\% & \textbf{53.33\%} & 46.66\% & 51.48\% & 48.12\% & 52.54\% & 50.77\% & 52.66\%\tabularnewline
\hline 
$\varLambda_{h}$ & 9.90\% & 10.29\% & \textbf{23.86\%} & 14.42\% & 16.98\% & 14.34\% & 17.59\% & 18.86\% & 17.79\%\tabularnewline
\hline 
$\varLambda_{m}$ & 18.10\% & 18.87\% & \textbf{42.95\%} & 26.26\% & 30.08\% & 26.02\% & 31.36\% & 34.01\% & 31.72\%\tabularnewline
\hline 
$time$ & 13.29 & 9.60 & 47.36 & 1.88 & \textbf{1.41} & 2.79 & 1.87 & 57.62 & 63.41\tabularnewline
\hline 
\end{tabular}
\end{table*}

\begin{figure*}
\begin{centering}
\includegraphics[width=15cm]{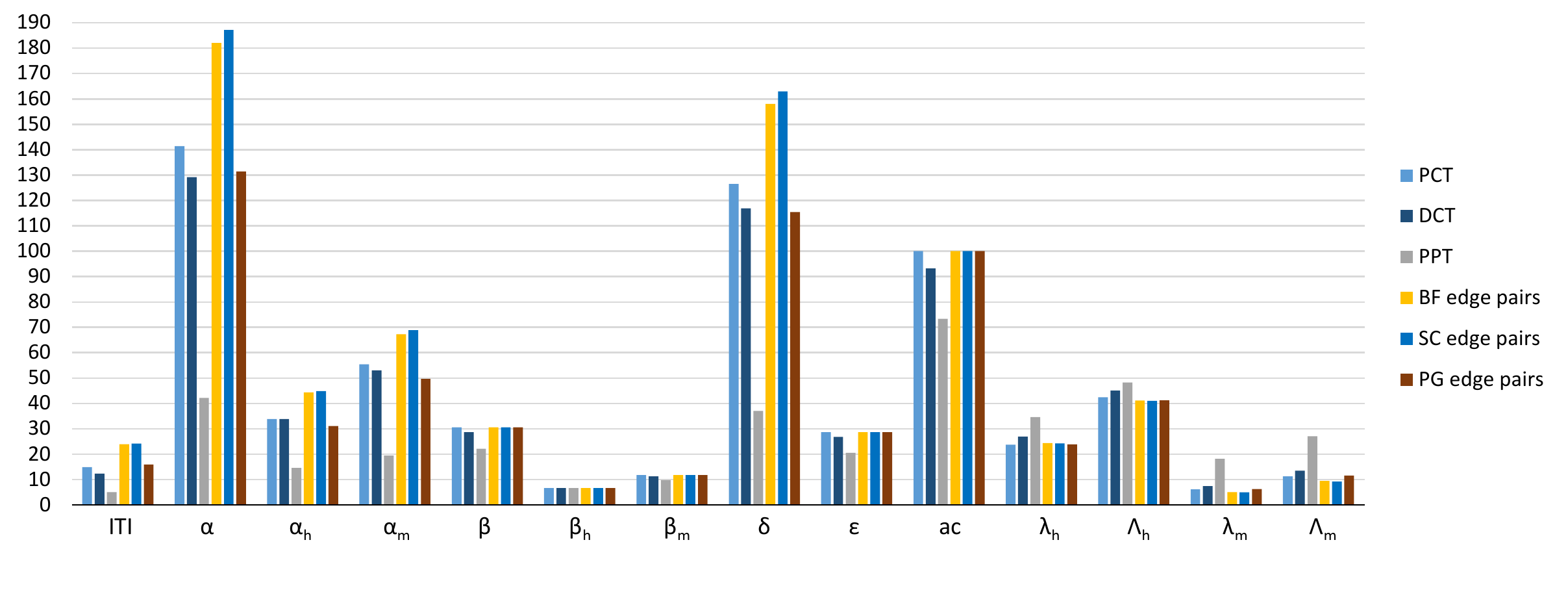}
\par\end{centering}
\caption{\label{fig:Comparison-of-algorithms TDL=00003D2 and PTL=00003Dhigh}Comparison
of algorithms for TDL=2 and PTL=high}
\end{figure*}

The average results of all 59 graphs for PCT, DCT(h), PPT(h), BF, SC and PG with $R$ created as edge pairs for $TDL=2$ and $PTL=high$ are presented in Table \ref{tab:results_TDL=00003D2_PTL=00003Dhigh} and a comparison of their average values are depicted in Figure \ref{fig:Comparison-of-algorithms TDL=00003D2 and PTL=00003Dhigh}.

\begin{table*}
\scriptsize

\caption{\label{tab:results_TDL=00003D2_PTL=00003Dhigh}Experiment results
for TDL=2, PTL=high. Detailed data for all graphs are available in Appendix C.}

\begin{singlespace}
\centering{}%
\begin{tabular}{|c|c|c|c|c|c|c|}
\hline 
Metric / $\mathcal{G}$  & \textbf{PCT} & \textbf{DCT(h)} & \textbf{PPT(h)} & \textbf{BF} edge pairs & \textbf{SC} edge pairs & \textbf{PG} edge pairs\tabularnewline
\hline 
\hline 
$\mid T\mid$ & 14.92 & 12.32 & \textbf{5.14} & 23.93 & 24.15 & 15.98\tabularnewline
\hline 
$\alpha$ & 141.41 & 129.19 & \textbf{42.19} & 182.03 & 187.14 & 131.39\tabularnewline
\hline 
$\alpha_{h}$ & 33.86 & 33.86 & \textbf{14.54} & 44.37 & 44.88 & 31.14\tabularnewline
\hline 
$\alpha_{m}$ & 55.39 & 52.95 & \textbf{19.53} & 67.27 & 68.90 & 49.66\tabularnewline
\hline 
$\beta$ & 30.59 & 28.63 & \textbf{22.12} & 30.59 & 30.59 & 30.59\tabularnewline
\hline 
$\beta_{h}$ & \textbf{6.71} & \textbf{6.71} & \textbf{6.71} & \textbf{6.71} & \textbf{6.71} & \textbf{6.71}\tabularnewline
\hline 
$\beta_{m}$ & 11.78 & 11.29 & \textbf{9.80} & 11.78 & 11.78 & 11.78\tabularnewline
\hline 
$\delta$ & 126.49 & 116.86 & \textbf{37.05} & 158.10 & 162.98 & 115.41\tabularnewline
\hline 
$\varepsilon$ & 28.68 & 26.88 & \textbf{20.59} & 28.68 & 28.68 & 28.68\tabularnewline
\hline 
$ac$  & 100\% & 93.24\% & \textbf{73.33\%} & 100\% & 100\% & 100\%\tabularnewline
\hline 
$\lambda_{h}$ & 23.79\% & 26.93\% & \textbf{34.63\%} & 24.44\% & 24.34\% & 23.85\%\tabularnewline
\hline 
$\lambda_{m}$ & 42.38\% & 45.12\% & \textbf{48.25\%} & 41.09\% & 41.04\% & 41.26\%\tabularnewline
\hline 
$\varLambda_{h}$ & 6.12\% & 7.46\% & \textbf{18.19\%} & 5.09\% & 4.93\% & 6.24\%\tabularnewline
\hline 
$\varLambda_{m}$ & 11.25\% & 13.49\% & \textbf{27.04\%} & 9.49\% & 9.20\% & 11.56\%\tabularnewline
\hline 
$time$ & 17.57 & 14.61 & 65.84 & \textbf{3.11} & 13.12 & 131.73\tabularnewline
\hline 
\end{tabular}
\end{singlespace}
\end{table*}

\begin{figure*}
\begin{centering}
\includegraphics[width=15cm]{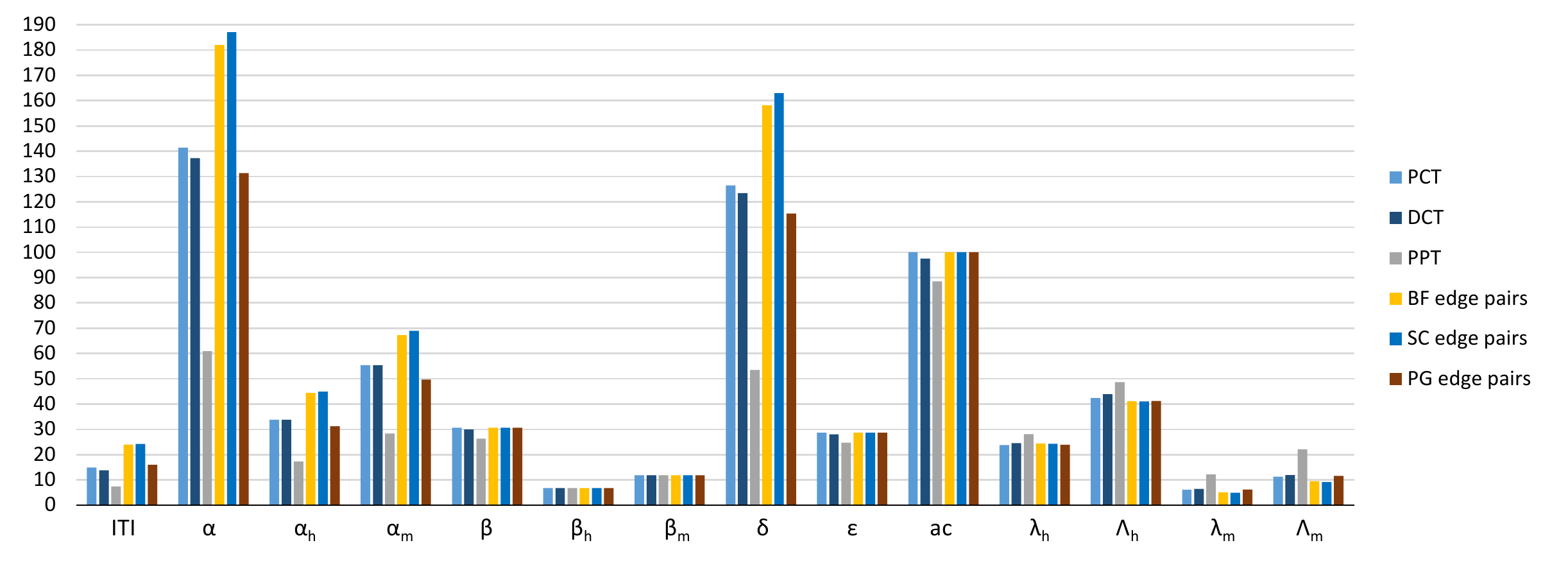}
\par\end{centering}
\caption{\label{fig:Comparison-of-algorithms for TDL=00003D2 and PTL=00003Dmedium}Comparison
of algorithms for TDL=2 and PTL=medium}
\end{figure*}

Finally, Table \ref{tab:results_TDL=00003D2_PTL=00003Dmedium} presents the averaged experimental results for all 59 graphs for the PCT, DCT(m), PPT(m), BF, SC and PG algorithms with $R$ for $TDL=2$ and $PTL=medium$. The results of the DCT(m) and PPT(m) algorithms in these test coverage criteria can be compared against PCT with $TDL=2$ and BF, SC and PG with $R$ created as edge pairs, acting as baselines when no prioritization is reflected in the test set. A comparison of the algorithms for this coverage criteria is presented in Figure \ref{fig:Comparison-of-algorithms for TDL=00003D2 and PTL=00003Dmedium}.

\begin{table*}
\scriptsize

\caption{\label{tab:results_TDL=00003D2_PTL=00003Dmedium}Experiment results
for TDL=2, PTL=medium. Detailed data for all graphs are available in Appendix D.}

\begin{singlespace}
\centering{}%
\begin{tabular}{|c|c|c|c|c|c|c|}
\hline 
Metric / $\mathcal{G}$  & \textbf{PCT} & \textbf{DCT(m)} & \textbf{PPT(m)} & \textbf{BF} edge pairs & \textbf{SC} edge pairs & \textbf{PG} edge pairs\tabularnewline
\hline 
\hline 
$\mid T\mid$ & 14.92 & 13.80 & \textbf{7.36} & 23.93 & 24.15 & 15.98\tabularnewline
\hline 
$\alpha$ & 141.41 & 137.25 & \textbf{60.85} & 182.03 & 187.14 & 131.39\tabularnewline
\hline 
$\alpha_{h}$ & 33.86 & 33.86 & \textbf{17.34} & 44.37 & 44.88 & 31.14\tabularnewline
\hline 
$\alpha_{m}$ & 55.39 & 55.39 & \textbf{28.36} & 67.27 & 68.90 & 49.66\tabularnewline
\hline 
$\beta$ & 30.59 & 29.88 & \textbf{26.36} & 30.59 & 30.59 & 30.59\tabularnewline
\hline 
$\beta_{h}$ & \textbf{6.71} & \textbf{6.71} & \textbf{6.71} & \textbf{6.71} & \textbf{6.71} & \textbf{6.71}\tabularnewline
\hline 
$\beta_{m}$ & \textbf{11.78} & \textbf{11.78} & \textbf{11.78} & \textbf{11.78} & \textbf{11.78} & \textbf{11.78}\tabularnewline
\hline 
$\delta$ & 126.49 & 123.46 & \textbf{53.49} & 158.10 & 162.98 & 115.41\tabularnewline
\hline 
$\varepsilon$ & 28.68 & 28.05 & \textbf{24.68} & 28.68 & 28.68 & 28.68\tabularnewline
\hline 
$ac$  & 100\% & 97.50\% & \textbf{88.51\%} & 100\% & 100\% & 100\%\tabularnewline
\hline 
$\lambda_{h}$ & 23.79\% & 24.53\% & \textbf{28.06\%} & 24.44\% & 24.34\% & 23.85\%\tabularnewline
\hline 
$\lambda_{m}$ & 42.38\% & 43.82\% & \textbf{48.61\%} & 41.09\% & 41.04\% & 41.26\%\tabularnewline
\hline 
$\varLambda_{h}$ & 6.12\% & 6.45\% & \textbf{12.09\%} & 5.09\% & 4.93\% & 6.24\%\tabularnewline
\hline 
$\varLambda_{m}$ & 11.25\% & 11.91\% & \textbf{22.08\%} & 9.49\% & 9.20\% & 11.56\%\tabularnewline
\hline 
$time$ & 17.57 & 15.33 & 77.15 & \textbf{3.11} & 13.12 & 131.73\tabularnewline
\hline 
\end{tabular}
\end{singlespace}
\end{table*}

In all four cases of measured test coverage (Tables \ref{tab:results_TDL=00003D1_PTL=00003Dhigh}, \ref{tab:results_TDL=00003D1_PTL=00003Dmedium}, \ref{tab:results_TDL=00003D2_PTL=00003Dhigh} and \ref{tab:results_TDL=00003D2_PTL=00003Dmedium}), the value of $\beta_{h}$ is equal for all of the measured algorithms. This selection of $\beta_{h}$ is fair and correct, as it represents a number of unique edges of priority $high$ covered by the test cases. For $PTL=medium$ (tables \ref{tab:results_TDL=00003D1_PTL=00003Dmedium} and \ref{tab:results_TDL=00003D2_PTL=00003Dmedium}), the value of $\beta_{m}$ is also equal for all compared algorithms. This result is also correct, as $\beta_{m}$ represents a number of unique edges of priority $high$ or $medium$ covered by the test cases. The other metrics indicate differences in the performance of the individual algorithms, which we discuss in the following subsection.

\subsection{\label{subsec:Discussion}Discussion}

Several issues can be observed from the experiments. First, the comparability issue of the individual algorithms is considered. For $TDL=1$ test coverage (or\textit{ All Edge Coverage}), we compared the PCT, DCT, PPT, BF, SC and PG algorithms for $PTL=high$ and $PTL=medium$. Moreover, for BF, SC and PG, we compared two alternative ways of creating test requirements $R$ can be created from $\mathcal{G}$: atomic and sequence conversion (specified in Table \ref{tab:Test_requirements_conversion_specification}). Regarding \textit{All Edge Coverage}, this approach is valid because all algorithms satisfy the \textit{All Edge Coverage} criteria. For BF, the test requirements are used to reflect the priorities of the individual edges. Thus, PCT serves as a baseline (as no prioritization of the edges is reflected there) and the DCT, PPT, BF, SC and PG algorithms can be compared in terms of the effectiveness with which they reflect the defined priorities.

For $TDL=2$ test coverage (or Edge-Pair coverage), we also compared the PCT, DCT, PPT, BF, SC and PG algorithms for $PTL=high$ and $PTL=medium$. Here, a situation differs in the way how BF, SC and PG algorithms is comparable with the other algorithms. In the BF, SC and PG algorithms, we used the test requirements to satisfy $TDL=2$ coverage. Thus, no option has been left to allow the algorithm to reflect the priorities defined in $\mathcal{G}$. Hence, for $TDL=2$, PCT, BF, SC and PG serve as baselines (as they reflect no prioritization of the edges), and the DCT and PPT algorithms can be compared in terms of the effectiveness with which they reflect the defined priorities.

For $TDL=1$, PPT outperforms the compared algorithms for the majority of the indicators. The only exception is $PTL=medium$ and $\lambda_{h}$, which is the ratio of the edges of priority $high$ and all edges contained in a test set $T$ (in percentage). In this case, SC and PG algorithms using sequence conversion of the test requirements performs better than PPT. 

Regarding the number of tests steps ($\alpha$), which can be considered as the main indicator related to the amount of work needed to exercise the produced test cases, for $PTL=high$, an average figure drops from 85.97 in case of baseline PPT (when no prioritization is used) to 78.00 in case of DCT(h). Obviously, the naive test set reduction strategy DCT does perform well in this point. Then, $\alpha$ drops to 29.20 in case of BF algorithm with sequence conversion of $R$, reflecting the priorities in the generation of the test cases systematically using $R$. Compared to DCT(h), this is a significat reduction. This reduction is further slightly improved by SC algorithm with sequence conversion of $R$ ($\alpha$ drops to 27.95) and further on very slightly by PG algorithm with sequence conversion of $R$ ($\alpha$ drops further to 27.42). However, the PPT algorithm reduces this average number further to 19.90, which is a 27.4\% difference compared to PG with sequence conversion of $R$. 

In case of $TDL=1$ and $PTL=medium$, the density of priority edges which should be considered in the test cases is higher. In this situation, PPT produced test cases with 31.27 steps in average, which is the most significant reduction compared to PG, where this figure was 42.86 for atomic conversion of test requirements. In this case, the difference of $\alpha$ between PPT and PG is 27\%.

Considering the number of tests steps $\alpha$, for $TDL=2$, a naive reduction of test cases simulated by DCT algorithm does not reduce the test set significantly when compared to PPT. In this case, PPT represents a considerable alternative for the situations when we try to achieve Edge-Pair coverage, and in parallel with that, to reflect priorities of edges in the model $\mathcal{G}$. Analyzing the data, this conclusion can be made for the $PTL=high$ as well as $PTL=medium$. 

When analysing the performance of PG algorithms for $TDL=2$, this algorithm produced more optimal $T$ without reflection of priorities in the model compared to PCT, PG and SC algorithms. This is documented by $\alpha$, $\alpha_{h}$, $\alpha_{m}$ and $\delta$ , whose values are better for PG than for PCT, PG and SC.

When discussing the effort needed to execute the produced test cases, another indicator should also be considered: the number of unique edges in the test cases ($\beta$). It is reasonable to assume that unique test steps will take more time to execute, as the testing situation is new for the tester. In test automation, this factor is even more significant, as we need to implement a new test script for a new unique test step. In this indicator, PPT also introduces a certain improvement; however, it is smaller than that of the total test steps ($\alpha$). Regarding the $\beta$ values for $TDL=1$ and $PTL=high$, the second best result following the PPT algorithm has been achieved by PG algorithm with sequence conversion of $R$. Compared to this algorithm, PPT further reduced $\beta$ by 5.5\%. For $PTL=medium$, the second best result has been also achieved by PG algorithm with sequence conversion of $R$, however improvement in $\beta$ further achieved by PPT can be considered insignificant (1.8\%). 

Regarding the execution time, for $TDL=1$, PPT is outperformed by BF and SC algorithms; the average execution time of BF and SC is approximately 15-25 times better than PPT average execution time (depends on $PTL$ ). The PPT is also outperformed by PCT algorithm, as the PCT average execution times is approximately 2.5 times better for $PTL=high$ and 3.5 times better for $PTL=medium$. The PPT overperformed PG algorithm and the difference in average execution time is approximately 30\% for $PTL=high$ and 20\% for $PTL=medium$. For $TDL=2$, the trend is the same with two differences: for this test coverage level, BF starts outperforming SC and difference between average execution time of PPT starts to be more significant than average execution time of PG (for $PTL=high$ this difference is approximately 50\% and for $PTL=medium$ approximately 40\%).

Considering the average execution times, the highest value for PPT is 77.15 milliseconds for $TDL=2$ and $PTL=medium$, which is the highest test coverage level examined in the experiments. When analyzing runtimes of PPT for individual problem instances for $TDL=2$ and $PTL=medium$ (refer to Appendix D), for the largest problem instances the runtime of PPT has not exceeded 0.6 seconds.

\section{\label{sec:Threats-to-Validity}Threats to Validity}

Although our goal was to design the experiments to be maximally transparent and objective, several concerns can be raised regarding the validity of the results. In this section, we list these issues and discuss countermeasures we have taken to minimize their impact.

The comparability of the PCT, DCT, PPT, BF, SC and PG algorithms in the experiments was discussed at the beginning of Section \ref{subsec:Discussion}. An issue related to this discussion is the objective creation of a set of test requirements $R$ from the SUT model $\mathcal{G}$. We solved this problem by measuring two possible alternatives of $R$ creation: atomic conversion and sequence conversion (specified in Table \ref{tab:Test_requirements_conversion_specification}). 

Another issue is that the PCT, DCT and PPT algorithms compose the test case as a sequence of $\mathcal{G}$ edges $a_{1},a{}_{2},..,a_{n-1}$, whereas BF, SC and PG compose the test case as a sequence of $G$ nodes $d_{1},d_{2},..,d_{n}$. In our implementation of the BF algorithm, we converted the test cases to edge sequences to make the BF results comparable with those of other algorithms. The same procedure was performed for SC and PG algorithms. The correctness of this conversion was verified thoroughly during the implementation tests and does not
affect the objectivity of the results. Except this conversion, we implemented the BF pseudo code without any changes or optimizations of the algorithm. To implement the  SC and PG algorithms, we used open-source code published by Offut et al. \cite{offuttools}, which eliminates possible biasws caused by our own implementation of the algorithms.

A question can be raised regarding the problem instances used in the experiments. However, the graphs used in the experiments were created manually to correspond to workflows of real software systems. In this process we used our knowledge and access to design documentation and source code for three real-world information systems: the medical information system Pluto, the customer relationship management system Global and the issue tracking system Mantis. Hence, the relevance of the problem instances is maintained. All three systems used in the experiments were selected to be tested by path-based techniques because of their practical suitability. The systems are workflow-based, provide multiple user roles and support processes of nontrivial complexity with a number of decision points and parallel branches of user scenarios triggered by these decision points, as documented by the complexity of the created process models used in the experiments. Regarding these properties, the systems serve as good cases for prioritized testing: coverage of the entire SUT by test cases of uniform intensity without any prioritization would, in a practical testing process, lead to either (1) a relatively small set of test cases with low test coverage (which would be realistic to execute with a given amount of time and resources but would possibly have lower effectiveness in detecting defects) or (2) a more extensive set of test cases with a higher probability to detect some defects (the execution of which is unfortunately very likely infeasible with a given amount of time and resources).

Regarding the measured execution times, for the BF, SC and PG algorithms, the generation time of $R$ from $\mathcal{G}$ edge priorities was not included in the measured execution time. We decided to exclude the test requirement conversion time to achieve more objective comparability of the algorithms with PPT. For the DCT algorithm, the PCT test cases were considered as an input to the test case reduction process, and the measured time covers only this reduction. Hence, the total time of DCT execution can be computed as a sum of the test case reduction time (i.e., the time presented for the DCT algorithm execution) and the PCT execution time. As this calculation is easy, we decided to present only the test case reduction time.

Regarding the performed experiments, we measured the performance of the algorithms on $TDL=1$ and $TDL=2$, which we consider as the test coverage being used in the majority of testing assignments for non-critical software systems \cite{van2013tmap,koomen2013tmap}. This test coverage is equivalent to \textit{All Edge Coverage} and \textit{Edge-pair Coverage} when we use an alternative terminology established in the area. Additional measurements can be performed for $TDL>2$, nevertheless, due to its limited usage \cite{van2013tmap,koomen2013tmap}, we decided to perform more intense experiments on the $TDL=1$ and $TDL=2$ coverage levels. 

\section{\label{sec:Related-Work}Related Work}

The current research on path-based testing techniques commonly uses a directed graph and a set of test requirements as an SUT model. Several test set optimization criteria can be defined (e.g., the minimal number of nodes, edges or paths satisfying the test requirements or coverage of test requirements by the produced test cases or alternatives \cite{li2009experimental}). Several algorithms have been explored to generate test cases \cite{dwarakanath2014minimum,li2012better}, such as Brute Force algorithm, Set-Covering Based Solution or a Matching-Based Prefix Graph Solution \cite{li2012better}. In addition, alternative strategies and algorithms for prioritizing the path-based test cases have been studied and developed. For instance, approaches based on neural network clustering \cite{gokcce2006coverage}, fuzzy clustering \cite{belli2007coverage} and the firefly optimization algorithm
\cite{panthi2015generating} have been implemented. In contrast to algorithms based on an SUT model and a set of test requirements only, in these techniques, information about the SUT internal structure is also used as an input to the prioritization.

The path-based techniques generally operate on an abstract level of an SUT model. These techniques can be applied to different levels of the SUT, where a process flow can be exercised. Three typical applications are: (1) testing of business processes and workflows related to functional end-to-end testing performed by testers (or an automated test) that exercise the SUT functions in a GUI \cite{bures2015pctgen}, (2) path testing on the code level \cite{yan2008efficient,li2009experimental} and (3) integration testing of larger sub-processes. 

In this type of integration testing, we focus on complex subprocesses with a process logic that can be modeled by a directed graph rather than close-API atomic integration tests, in which we focus rather on defining the API parameters to call and assert the expected results.

For path testing on the code level, Control Flow Graphs, conceptually similar to $G$, were used by \cite{yan2008efficient} as SUT models. The related work in this domain starts to overlap with data-flow testing, which is more oriented toward checking the consistency of the data processed inside an algorithm \cite{chaim2013efficient,denaro2014right,denaro2015dynamic,su2017survey}.

The test requirements concept is used as a common abstraction to determine which parts of the SUT model are prioritized for testing and have to be covered by the test cases \cite{yoo2012regression}. In addition to this purpose, the test requirements can also be used to specify a general test coverage criteria, such as \textit{Edge-pair Coverage} for instance, which can be seen as a limitation of this concept. For situations in which test requirements are used to define general test coverage, the requirements cannot be used again to specify priority parts of the SUT model (unless we perform a special determination of test requirements which would reflect the both aspects, which is a process, which is feasible to be done by a special dedicated algorithm created for this purpose, but hardly feasible manually in real-life praxis of test designers). 

In contrast to the test requirement concept, which can be practically used to set only one level of priority, more levels are used for prioritization in the common software engineering and test management practice \cite{achimugu2014systematic,van2013tmap}. However, this issue can be solved by transforming priority levels into a set of test requirements.

Due to the large volume of testing activities dedicated to process-oriented testing, further evolution of path-based techniques is a relevant topic. For the reasons we have summarized in Section \ref{subsec:Motivation-for-an}, the approach we propose in this paper represents an alternative to the previous research achievements in path-based testing methods.

\section{Conclusion}

In this paper, we proposed the Prioritized Process Test (PPT) algorithm in a strategy that generates path-based test cases from the SUT model. A weighted directed multigraph is used as a model for the SUT. The weights are the priorities of the SUT functions, which must be reflected in the generated test set $T$. This concept is an alternative to the currently established approach \cite{dwarakanath2014minimum,li2012better}, in which a plain directed graph is used as an SUT model and the priorities to test can be expressed by a set of test requirements (i.e., paths in the graph, which must be present in the test cases). In addition, the test requirements can also be used to determine the test coverage level of the test set.

At this point, the test requirement concept can reach a certain limit when we need to determine the test coverage level of the test cases and, in parallel, specify the priority of SUT functions, which should be reflected in the test set $T$. This limit can be demonstrated for $TDL=2$ (or Edge-pair) coverage.

The PPT algorithm allows both the test coverage level and the priorities to be specified in parallel, as it uses two concurrent test coverage criteria: Test Depth Level (TDL) and Prioritized Test Level (PTL). As the experimental results have shown, PPT represents an alternative for situations in which we try to achieve \textit{Edge-pair Coverage} and, in parallel, reflect the priorities of functions or actions in the SUT model. Additionally, for \textit{All Edge Coverage} ($TDL=1$), the algorithm produced more optimal test set $T$ than the BF, SC and PG algorithms for a number of aspects.

Several optimization criteria can be discussed when analyzing the test set. In this paper, we provided comprehensive data for nine metrics based on test case properties and another five metrics based on various indicators computed for these properties. The most important issue is the minimization of the test set. Considering the total number of test steps, for $PTL=high$ the PPT algorithm reduced this number by 27.4\% on average compared to the best result achieved by the alternatives (namely PG with sequence conversion of $R$ ). For $PTL=medium$ the PPT algorithm reduced the total number of test steps by 27\% on average compared to the best result achieved by the alternatives (PG with atomic conversion of $R$ in this case). When considering the number of unique test steps, the PPT algorithm reduced this number by 5.5\% compared to PG algorithm with sequence conversion of $R$ (which yielded the second best result following PPT algorithm).

Regarding future work, we are currently optimizing the implementation of the PPT algorithm on the Oxygen platform. In parallel, we are adapting PPT to accept a set of defined preconditions in the input and reflect these preconditions in the generated test set. Currently, we are considering two types of preconditions: (1) if an edge or a node is visited ina test case, another set of edges or nodes must also be visited, and (2) if an edge or a node is visited in a test case, another set of edges or nodes must not be avoided in the test case. These preconditions are inspired by our discussions with testing industry practitioners and model several real-life situations in the testing of software and IoT systems, such as the prerequisites required to create test data during the process tests, the handling of different user access roles in the SUT, the possible run-time optimization of the test set based on collected data on defects present in the SUT in previous iterations of the software development process, or the testing of the limited network connectivity in IoT systems from a process viewpoint.

\section*{Acknowledgments}

This research is conducted as a part of the project TACR TH02010296 Quality Assurance System for Internet of Things Technology.

\bibliographystyle{ws-ijseke}
\bibliography{sample.bib}

\begin{thebibliography}{10}

\bibitem{offutt2008introduction}
P.~Ammann and J.~Offutt, {\em Introduction to software testing} (Cambridge
  University Press, 2016).

\bibitem{myers2011art}
G.~J. Myers, C.~Sandler and T.~Badgett, {\em The art of software testing} (John
  Wiley \& Sons, 2011).

\bibitem{Eldh:2006}
S.~Eldh, H.~Hansson, S.~Punnekkat, A.~Pettersson and D.~Sundmark, A framework
  for comparing efficiency, effectiveness and applicability of software testing
  techniques, in {\em Testing: Academic Industrial Conference - Practice And
  Research Techniques (TAIC PART'06)\/}, Aug 2006, pp. 159--170.

\bibitem{koomen2013tmap}
T.~Koomen, B.~Broekman, L.~van~der Aalst and M.~Vroon, {\em TMap next: for
  result-driven testing} (Uitgeverij kleine Uil, 2013).

\bibitem{utting2010practical}
M.~Utting and B.~Legeard, {\em Practical model-based testing: a tools approach}
  (Elsevier, 2010).

\bibitem{gupta2008approach}
A.~Gupta and P.~Jalote, An approach for experimentally evaluating effectiveness
  and efficiency of coverage criteria for software testing, {\em International
  Journal on Software Tools for Technology Transfer} {\bf 10}(2)  (2008)
  145--160.

\bibitem{schieferdecker2012model}
I.~Schieferdecker, Model-based testing, {\em IEEE software} {\bf 29}(1)  (2012)
  p.~14.

\bibitem{utting2012taxonomy}
M.~Utting, A.~Pretschner and B.~Legeard, A taxonomy of model-based testing
  approaches, {\em Software Testing, Verification and Reliability} {\bf 22}(5)
  (2012)  297--312.

\bibitem{budgen2011empirical}
D.~Budgen, A.~J. Burn, O.~P. Brereton, B.~A. Kitchenham and R.~Pretorius,
  Empirical evidence about the uml: a systematic literature review, {\em
  Software: Practice and Experience} {\bf 41}(4)  (2011)  363--392.

\bibitem{rumpe2016modeling}
B.~Rumpe, Modeling with uml, {\em Language, Concepts, Methods. Springer
  International} {\bf 4}  (2016).

\bibitem{wieczorek2009applying}
S.~Wieczorek, V.~Kozyura, A.~Roth, M.~Leuschel, J.~Bendisposto, D.~Plagge and
  I.~Schieferdecker, Applying model checking to generate model-based
  integration tests from choreography models, in {\em Testing of Software and
  Communication Systems\/},  (Springer, 2009) pp. 179--194.

\bibitem{kalenkova2017process}
A.~A. Kalenkova, W.~M. van~der Aalst, I.~A. Lomazova and V.~A. Rubin, Process
  mining using bpmn: relating event logs and process models, {\em Software \&
  Systems Modeling} {\bf 16}(4)  (2017)  1019--1048.

\bibitem{mishra2015graph}
A.~Mishra and A.~Sureka, A graph processing based approach for automatic
  detection of semantic inconsistency between bpmn process model and sbvr
  rules, in {\em International Conference on Mining Intelligence and Knowledge
  Exploration\/},  Springer2015, pp. 115--129.

\bibitem{yoo2012regression}
S.~Yoo and M.~Harman, Regression testing minimization, selection and
  prioritization: a survey, {\em Software Testing, Verification and
  Reliability} {\bf 22}(2)  (2012)  67--120.

\bibitem{dwarakanath2014minimum}
A.~Dwarakanath and A.~Jankiti, Minimum number of test paths for prime path and
  other structural coverage criteria, in {\em IFIP International Conference on
  Testing Software and Systems\/},  Springer2014, pp. 63--79.

\bibitem{li2012better}
N.~Li, F.~Li and J.~Offutt, Better algorithms to minimize the cost of test
  paths, in {\em Software Testing, Verification and Validation (ICST), 2012
  IEEE Fifth International Conference on\/},  IEEE2012, pp. 280--289.

\bibitem{gokcce2006coverage}
N.~G{\"o}k{\c{c}}e, M.~Eminov and F.~Belli, Coverage-based, prioritized testing
  using neural network clustering, in {\em International Symposium on Computer
  and Information Sciences\/},  Springer2006, pp. 1060--1071.

\bibitem{belli2007coverage}
F.~Belli, M.~Eminov and N.~G{\"o}k{\c{c}}e, Coverage-oriented, prioritized
  testing--a fuzzy clustering approach and case study, in {\em Latin-American
  Symposium on Dependable Computing\/},  Springer2007, pp. 95--110.

\bibitem{panthi2015generating}
V.~Panthi and D.~Mohapatra, Generating prioritized test sequences using firefly
  optimization technique, in {\em Computational Intelligence in Data
  Mining-Volume 2\/},  (Springer, 2015) pp. 627--635.

\bibitem{achimugu2014systematic}
P.~Achimugu, A.~Selamat, R.~Ibrahim and M.~N. Mahrin, A systematic literature
  review of software requirements prioritization research, {\em Information and
  software technology} {\bf 56}(6)  (2014)  568--585.

\bibitem{van2013tmap}
L.~van~der Aalst, E.~Roodenrijs, J.~Vink and R.~Baarda, {\em TMap NEXT:
  business driven test management} (Uitgeverij kleine Uil, 2013).

\bibitem{offuttoolsapplication}
P.~Ammann and J.~Offutt, Graph coverage web application,
  http://cs.gmu.edu:8080/offutt/coverage/graphcoverage  (2017).

\bibitem{bures2017prioritized}
M.~Bures, T.~Cerny and M.~Klima, Prioritized process test: More efficiency in
  testing of business processes and workflows, in {\em International Conference
  on Information Science and Applications\/},  Springer2017, pp. 585--593.

\bibitem{bures2015pctgen}
M.~Bures, Pctgen: automated generation of test cases for application workflows,
  in {\em New Contributions in Information Systems and Technologies\/},
  (Springer, 2015) pp. 789--794.

\bibitem{offuttools}
P.~A. Jeff~Offutt and N.~Li, Source code of coverage web applications,
  http://cs.gmu.edu/~offutt/softwaretest/coverage-source/  (2015).

\bibitem{li2009experimental}
N.~Li, U.~Praphamontripong and J.~Offutt, An experimental comparison of four
  unit test criteria: Mutation, edge-pair, all-uses and prime path coverage, in
  {\em Software Testing, Verification and Validation Workshops, 2009. ICSTW'09.
  International Conference on\/},  IEEE2009, pp. 220--229.

\bibitem{yan2008efficient}
J.~Yan and J.~Zhang, An efficient method to generate feasible paths for basis
  path testing, {\em Information Processing Letters} {\bf 107}(3-4)  (2008)
  87--92.

\bibitem{chaim2013efficient}
M.~L. Chaim and R.~P.~A. De~Araujo, An efficient bitwise algorithm for
  intra-procedural data-flow testing coverage, {\em Information Processing
  Letters} {\bf 113}(8)  (2013)  293--300.

\bibitem{denaro2014right}
G.~Denaro, M.~Pezz{\`e} and M.~Vivanti, On the right objectives of data flow
  testing, in {\em Software Testing, Verification and Validation (ICST), 2014
  IEEE Seventh International Conference on\/},  IEEE2014, pp. 71--80.

\bibitem{denaro2015dynamic}
G.~Denaro, A.~Margara, M.~Pezze and M.~Vivanti, Dynamic data flow testing of
  object oriented systems, in {\em Proceedings of the 37th International
  Conference on Software Engineering-Volume 1\/},  IEEE Press2015, pp.
  947--958.

\bibitem{su2017survey}
T.~Su, K.~Wu, W.~Miao, G.~Pu, J.~He, Y.~Chen and Z.~Su, A survey on data-flow
  testing, {\em ACM Computing Surveys (CSUR)} {\bf 50}(1)  (2017) p.~5.

\end{thebibliography}

\section*{Appendices}

\subsection*{Appendix A }

\href{http://still.felk.cvut.cz/download/Appendix\_A.pdf}{http://still.felk.cvut.cz/download/Appendix\_A.pdf}

\subsection*{Appendix B}

\href{http://still.felk.cvut.cz/download/Appendix\_B.pdf}{http://still.felk.cvut.cz/download/Appendix\_B.pdf}

\subsection*{Appendix C }

\href{http://still.felk.cvut.cz/download/Appendix\_C.pdf}{http://still.felk.cvut.cz/download/Appendix\_C.pdf}

\subsection*{Appendix D}

\href{http://still.felk.cvut.cz/download/Appendix\_D.pdf}{http://still.felk.cvut.cz/download/Appendix\_D.pdf}

\end{document}